\documentstyle[12pt,aasms4]{article}
\begin{document}

\title{ LOPSIDEDNESS IN  DWARF IRREGULAR GALAXIES}
\vspace{1.0 truecm}
\vspace{1.0 truecm}
\author{Ana B. Heller,  Noah  Brosch, Elchanan Almoznino}
\affil{The Wise Observatory and 
the School of Physics and Astronomy \\ Tel Aviv University, Tel Aviv 69978,  
Israel.}
\vspace{0.5 truecm}
\author{Liese van Zee}
\affil{Herzberg Institute of Astrophysics\\
5071 W.  Saanich Rd., Victoria BC V8X 4M6, Canada.}
\vspace{0.5 truecm}
 \and
\author{John J. Salzer}
\affil{Astronomy Department, Wesleyan University,\\
 Middletown, CT 06459-0123.}
\vspace{1.0 truecm}
\vspace{1.0 truecm}
\vspace{1.0 truecm}

\begin{abstract}
We quantify the amplitude of the lopsidedness,
 the azimuthal  angular asymmetry index, and the 
concentration of star forming regions,
 as represented by the distribution of the  H$\alpha$ 
emission,  in a sample of 78 late-type irregular galaxies.
We bin the observed galaxies in  two groups
representing blue compact galaxies (BCDs) and low surface
brightness dwarf galaxies (LSBs). 
The light distribution is analysed with a novel algorithm, 
which allows detection of details in the light distribution pattern.
 We find that while the asymmetry
of the underlying continuum light,
representing the older stellar generations,
 is relatively small, the H$\alpha$ emission is
very asymmetric and is  correlated in  position angle with the continuum
 light.  We show that the 
concentration of  continuum light is correlated with the H$\alpha$
concentration; this implies that the young star formation has the same spatial
properties as the older stellar populations, but that these  properties are
 more strongly expressed
 by the young stars.
 We test a model of random star formation over the extent of a galaxy 
 by simulating HII regions in  artificial dwarf  galaxies.
 A galaxy is traced  by assuming   red star clusters
distributed on an  underlying exponential disk of radius
 twice the scale length. The disk is allowed to
change in apparent magnitude, scale radius, position angle, 
 and ellipticity.  
 We  compare the  asymmetry-concentration distribution predicted by
 the simulations with the real observed distribution;
 we find that only  LSBs match the  distribution predicted by the model. 
The reason is that, independently  of the number of HII regions,
  LSBs  show no particular location of HII regions, whereas
BCDs show current star formation activity restricted very much 
 to the central parts  of the galaxies. 
 A consideration of the properties
of the continuum light leads to the conclusion
that   most of  LSBs can be approximated by exponential disks of radius 
twice their scale lengths; BCDs
 call, however, for much more concentrated underlying systems,
 with smaller scale lengths than  assumed in  the simulations. 
The implication is  that  random star formation over the full extent
of a galaxy may be generated in LSB dwarf-irregular galaxies
 but not in BCD galaxies.  
\end{abstract}
\vspace{1cm}

\vspace{1.2 truecm}
\section{Introduction}
  One of the least understood aspects of galaxy evolution is the onset of
  lopsidedness
  in the gaseous and stellar distributions of disk galaxies.  Recent models
  of disk galaxies suggest that the presence of a dominant dark halo can both
  produce and help sustain asymmetries in the gaseous and stellar components.
  For instance,  a lopsided  gravitational potential of a 
 dark matter halo  can
   produce an asymmetric galaxy as the gas surface density responds
  to the overall asymmetry   (Jog  1997). Alternatively,
  a symmetric
  dark matter halo can produce an asymmetric galaxy if  the disk orbits
  off--center
  of the overall potential (Levine \& Sparke 1998). In addition to  these 
  ``intrinsic"
  models, the environment and merger history of a galaxy can  affect its
  present appearance. For example, recent dynamical simulations of the effect of
  an infalling satellite  indicate that tidal interactions are yet another
  mechanism by
  which asymmetric galaxies can be formed  
  (Walker, Mihos,  \& Hernquist 1996; Zaritsky \& Rix 1997).
   
 However, all the above models were designed to account
 for asymmetries observed  in massive spiral galaxies. 
  The question of asymmetries in dwarf irregular galaxies~(dIs) may demand
  a different approach, as dIs lack  spiral density waves and tidal
  shear forces  that contribute to  induce gas instabilities.
  In dIs, the gas appears to be close to stability
throughout the disk, even though star formation 
   is occurring (Hunter~{\it et al.}~1998).
    Moreover, there is growing evidence (Mihos, McGaugh \& de Blok 1997) that
   low-surface-brightness (LSB) disks are reasonably stable 
   and remain structurally intact during  tidal encounters. 
  In addition, various tests  of 
     Virgo dIs favor internal over external mechanisms of star
  formation
  (Heller~{\it et al.}~1998), with the implicit conclusion that
asymmetries may also form through internal mechanisms. 

  We conclude that, while theoretical arguments, such as
  the  presence of a  dominant
   dark matter halo potential in dIs, may  contribute 
   to the long-persistence of the asymmetries,
    other processes  of star formation, such as
   the Stochastic Self-Propagating Star Formation  (SSPSF, 
  Gerola, Seiden \& Schulman 1980),  or alternatively random gas compression
  from turbulence, or random collisions of ISM clouds (Larson 1986; Elmegreen 1998) 
  may be the dominant regulators of the star formation in  dIs. 
 This conclusion is supported by   recent star formation histories (SFH)
   of Sextans A and GR 8    derived from  HST observations. 
   A series of chronological  frames
  showing  the spatial distribution of blue HeB stars
indicate that, chronologically,  the star formation  activity is propagating around  in 
 these galaxies with typical sizes of $\sim$100 pc and lifetimes
  of order 100 Myr (Dohm-Palmer~{\it et al.}~1997, 1998a,~b). 
   The question  remains as to whether   random mechanisms may introduce
  temporary
   asymmetries  in the stellar and gaseous components of low-mass systems. 
  We  address this issue by
 (a)  analyzing  the light  distribution in deep narrow-band H$\alpha$
   and continuum
   images  of a large sample of star forming  dIs,
   (b) developing   a new impartial algorithm   to compute
   the lopsidedness of star-forming regions and simultaneously compare it to the
  distribution of the stellar component,
    (c)~constructing  1000 model  galaxies  and
   showing that it is the ``discrete" behavior of 
  random star forming regions  that produce the asymmetric 
structure observed in most of the dIs.

  The plan of the paper is as follows:  we first describe the sample
  of galaxies, which  is a collection of objects
     with  previously published observations. The analysis method is
  described next, then the  results are presented. Finally, we 
   describe the simulations performed to understand the observational results and
  their implications.

\section{The sample}

 The galaxy sample studied here    consists of 78   dIs observed
  with the
Wise Observatory (WO) 1.0 m telescope or with
 the Kitt Peak  National Observatory (KPNO) 0.9 m telescope.
  The galaxies are
 classified  in the original publications as dIs, with  absolute
blue magnitudes $<$ -18,  and are smaller than 2 arcmin. The only
restrictions to the inclusion in the sample are the availability of CCD
   H$\alpha$  images with detected HII regions,
  and \(v_{\odot}  \leq 3,000\)  km sec$^{-1}$ (except for UM408, which has 
 \(v_{\odot} = 3,492\)~km~sec$^{-1}$).
 All the selected galaxies appear to be isolated, with the exception of 
 objects marked  with an asterisk in Table~\ref{T1abc}.
   Representative H$\alpha$   images, catalog references,
 and extensive additional details,  can be found in
 van Zee  {\it et al.}~(1997a, b, c),   
Almoznino \& Brosch~(1998), 
 Heller  {\it et al.}~(1999, 2000), and Norton \& Salzer (2000).

In order to test  for  dependence of the SFR on the lopsidedness
 we divided the sample in two sub-groups.  The first, called here BCD,  is
 represented by  33 blue compact dwarf galaxies 
(classified  morphologically as BCD or anything+BCD:
 references 5, 6,
 and 7 in Table~\ref{T1abc}). These are   galaxies whose 
optical light output is  often dominated
by the  strong  starburst component.
The second group,  called LSB,   is represented by 45
  low surface brightness dwarf galaxies (references 1, 2,  3, and 4).  This group
 includes dIs, primarily from standard catalogues, which are gas-rich  and,
in general, have     central surface brightness fainter
 than  $\sim~23.0$~mag ~arcsec$^{-2}$.
  Some of the  more luminous LSB galaxies
  show evidence of spiral features and may belong to the ``dwarf spiral'' class
(UGC numbers 191, 634, 3050, 3174, 4660, 5716, 
 7178,  9762, 10281, and  11820). 
 The   typical  SFR  for the LSB group is
 $\sim  7 \times 10^{-3}$~M$_{\odot}$~yr$^{-1}$; this is,
on average, one order of magnitude weaker than for BCD objects,  
 although there
is overlap in SFR between the brighter LSBs and the fainter BCDs.

\section {Analysis and results}
\subsection{Method}
In general, the lopsidedness of a galaxy is measured on a broad-band image 
 (usually in the red).
Some authors (Zaritsky \& Rix 1997; Rudnick \& Rix 1998)  use the ratio of
 the m=1 to m=0 Fourier amplitudes of the image as a quantitative 
measure of lopsidedness
in early-type disk galaxies. Others take a
 more direct approach of comparing the integrated light
 within specified regions of the galaxy. 
 For instance, Kornreich {\it et al.}~(1998) compare the
relative fluxes within trapezoidal sectors arranged symmetrically about the
galaxy's center of light.
Similarly,  Abraham {\it et al.}~(1996) define the rotational 
asymmetry parameter as
 half the ratio of the absolute value of the  
  difference between the original galaxy image and the image rotated 
 by a half-turn
 about its center, to the original image. 
The rotational asymmetry parameter, together with
 the central concentration of the emitted flux, has proven
 to be an important tool  to
    extend  the  morphological classification of the galaxies
 from the nearby Universe to high redshifts.  Likewise,
color-asymmetry diagrams, when combined with  information
about the axial ratio, can be used  to disentangle interacting
 galaxies from non-interacting, face-on systems at high redshift
    (Conselice 1997;  Conselice \& Bershady 1999).

We have developed a new  method to evaluate 
the variation of asymmetry  with azimuthal position angle, 
and  also the concentration of the star forming regions 
and of the general stellar distribution. 
In our method,  the  H$\alpha$ line emission represents
the distribution of the recently formed massive stars  that are
 younger than a few tens of Myrs, while  the red
continuum emission represents the   distribution  of the integrated
stellar populations. This latter component may be contaminated
by nebular continuum emission from HII regions,
 but nebular emission amounts to only 30$\%$ of the light
during the first few Myrs of a starburst, and becomes negligible
as soon as the first red supergiants appear, at about 10$^{6.9}$ yr
 (Leitherer \& Heckman 1995). We  investigate both the
asymmetry and concentration properties of these two
components, as well as look for correlations between them.

 For most  galaxies in the sample, 
 two red, narrow-band images were used:  one centered on the
rest-frame  H$\alpha$ line ($H\alpha_{on}$) and the other 
 sampling  the continuum 
($Cont$) region  near  H$\alpha$.
 For some of the LSBs, the narrow-band continuum  images 
 were no longer available at the time of the present analysis.  
 For these systems (marked ``+'' in Table~\ref{T1abc}),
   sky-subtracted B broad-band images were only
 used to trace the ellipse contour of the galaxies (see below)
 but are not included in the  statistical results for $Cont$.
 In all  cases, the sky background was subtracted in each
band  and the net-H$\alpha$ ($H\alpha$) images were derived by subtracting
 $Cont$ from  $H\alpha_{on}$ images with proper scaling. Details are given
 in Heller~{\it et al.} (1999).

Due to  the lack of obvious central concentration and the irregular
shape of the galaxies, we performed different fits  of elliptical isophotes,
allowing the  ellipticity and the position angle to vary, and fitting
out to  25 mag arcsec$^{-2}$ on the continuum image. 
The convergence criterion for the final  parameters was set
when the outer isophote retained the position angle and ellipticity
 of the ellipse traced at half the major axis.
 From then on, the position angle,
the ellipticity, and  the extent of the galaxy were held fixed, and 
the outer ellipse contour  was transposed to the H$\alpha$ images. 
 For those objects  without calibrated images, 
(refs. 2,  5,  and 6 in Table~\ref{T1abc}),
 the  outermost isophote  was adopted at the level  where
  the mean  intensity reached the  sky fluctuations.  This choice
 presumably  depends on the depth of the exposure and on
  systematic errors in the subtraction of the sky background.
 The reduction was done with IRAF\footnote{IRAF is distributed by 
the National Optical Astronomy Observatories.}
 and the fitted  ellipse  parameters  are listed in Table~\ref{T1abc}.

\begin{figure}{}
\figurenum{1.1}
\epsscale{0.90}  
\plotone{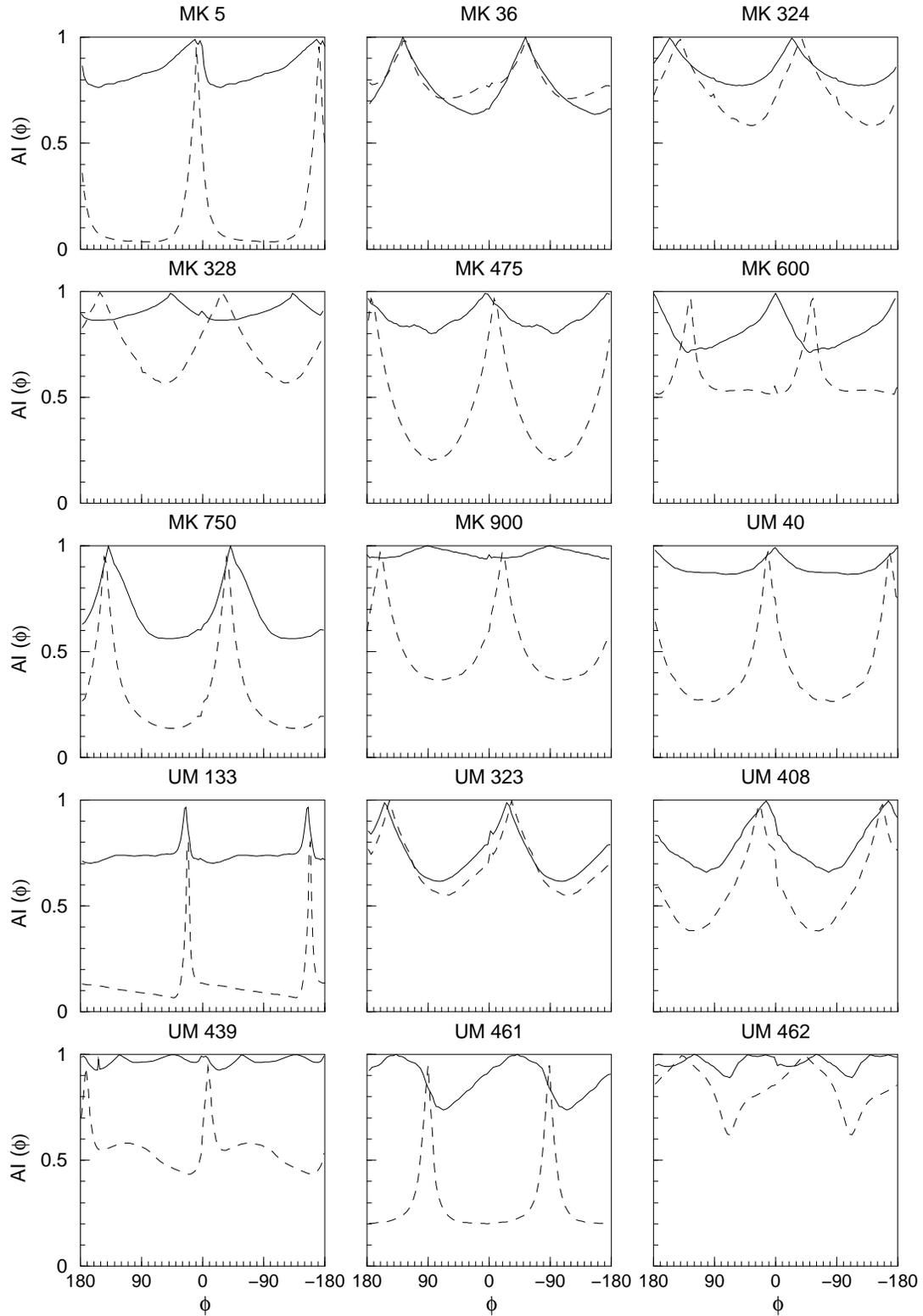}
\figcaption{BCDs: Variation of AI vs. azimuthal angle
                   ($\Phi$) measured anti-clockwise from North.
          ${Cont}$ is  plotted with a solid line and H$\alpha$ 
        is plotted with a dashed line. \label{S1}}
\end{figure}

\begin{figure}{}
\figurenum{1.2}
\epsscale{0.90}  
\plotone{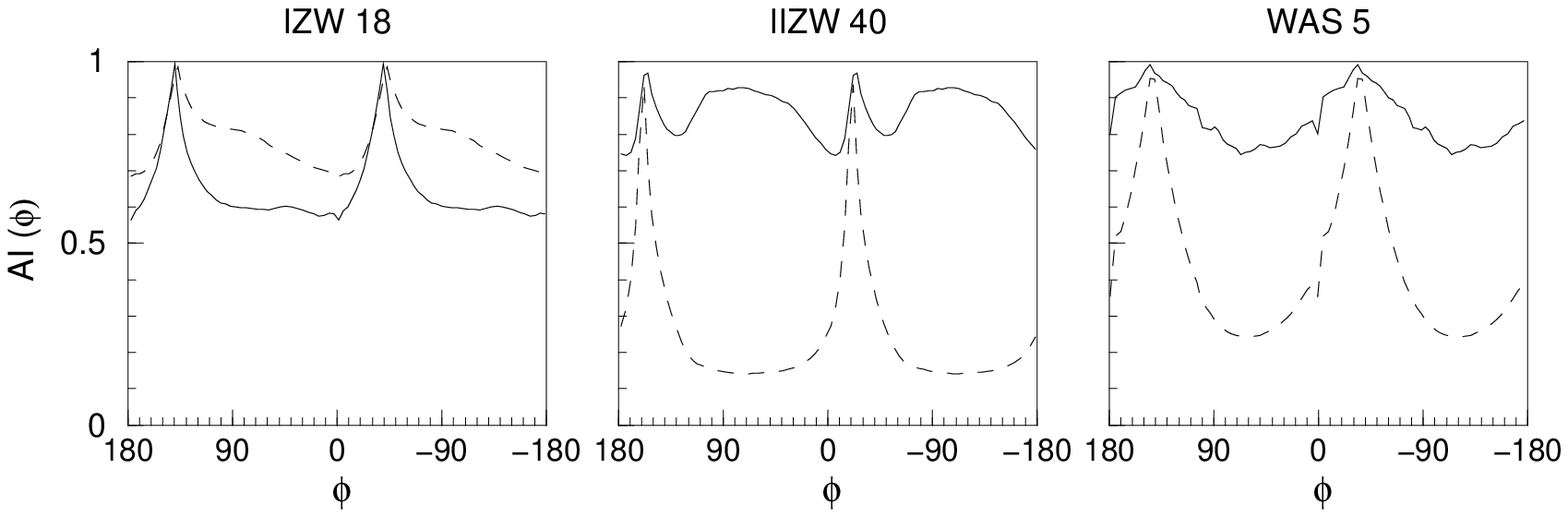}
\figcaption{Same as Fig.\ref{S1}. \label{S2}}
\end{figure}

\begin{figure}{}
\figurenum{1.3}
\epsscale{0.90}  
\plotone{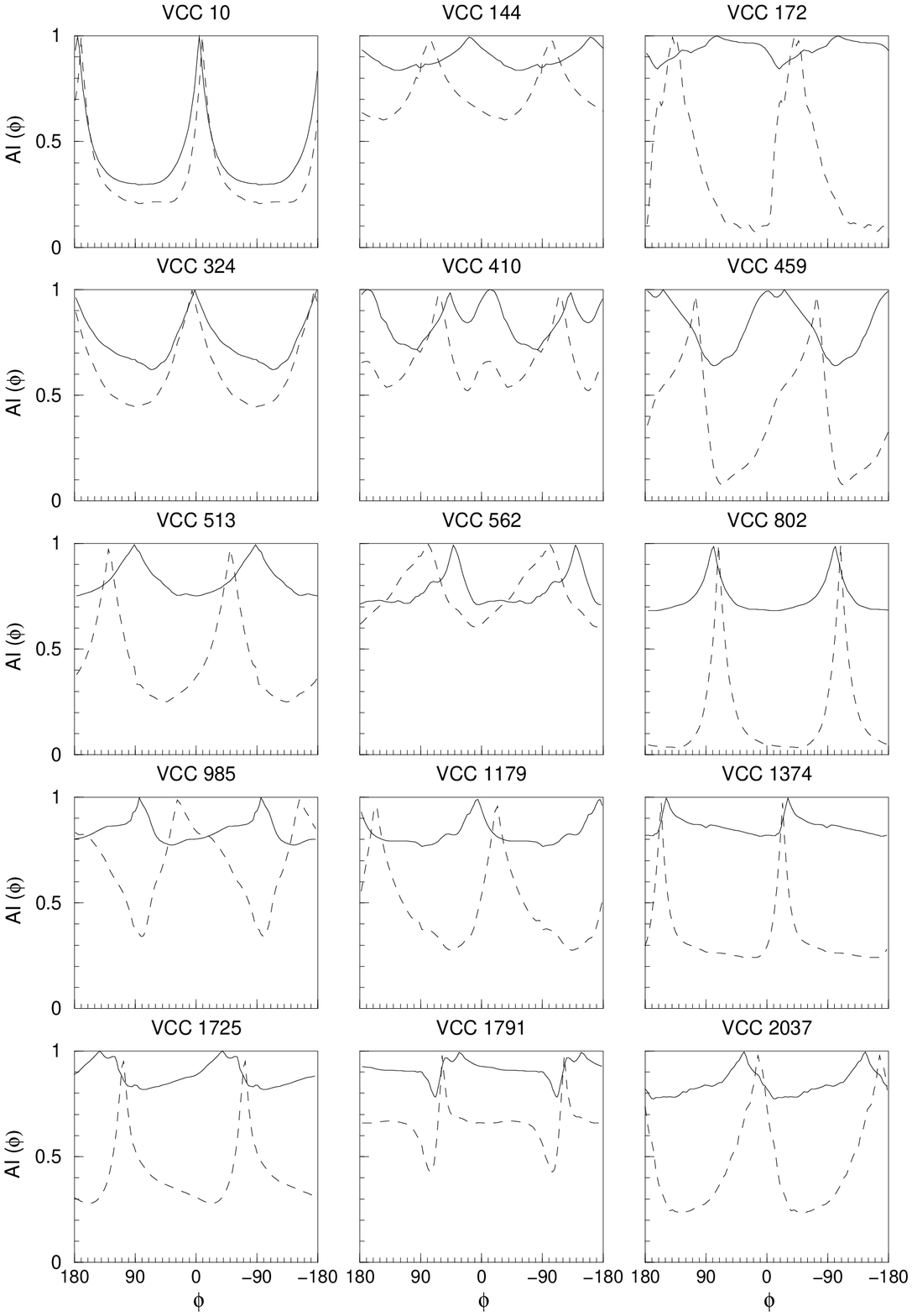}
\figcaption{Same as Fig. ~\ref{S1}. \label{S3}}
\end{figure}

\begin{figure}{}
\figurenum{2.1}
\epsscale{0.90}  
\plotone{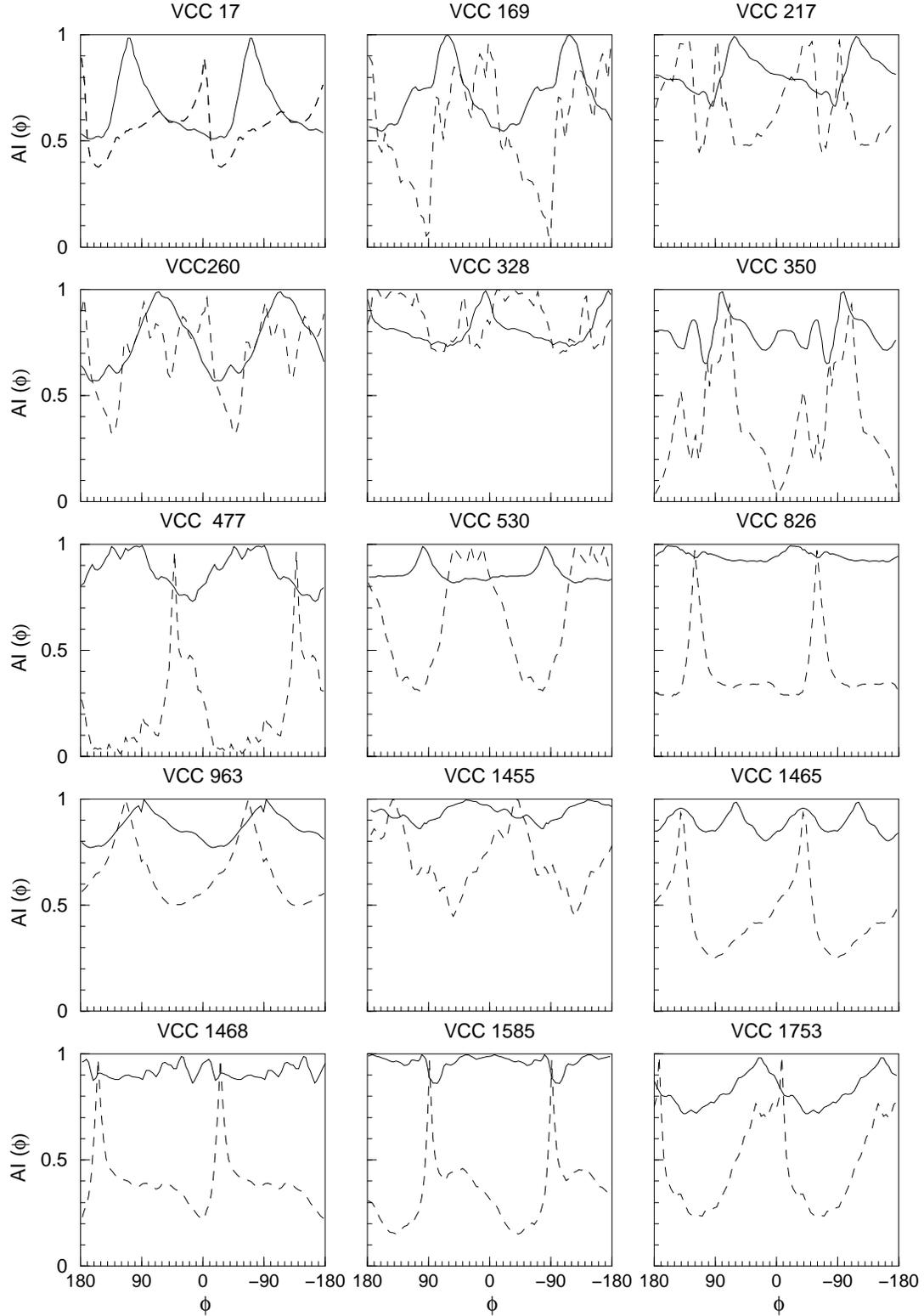}
\figcaption{LSBs: Variation of AI vs. azimuthal angle
                   ($\Phi$) for LSBs  with
                     narrow-band  $H\alpha$     images for the continuum. 
          ${Cont}$ is  plotted with a solid line and H$\alpha$ 
        is plotted with a dashed line. \label{S4}}
\end{figure}

\begin{figure}{}
\figurenum{2.2}
\epsscale{0.90}  
\plotone{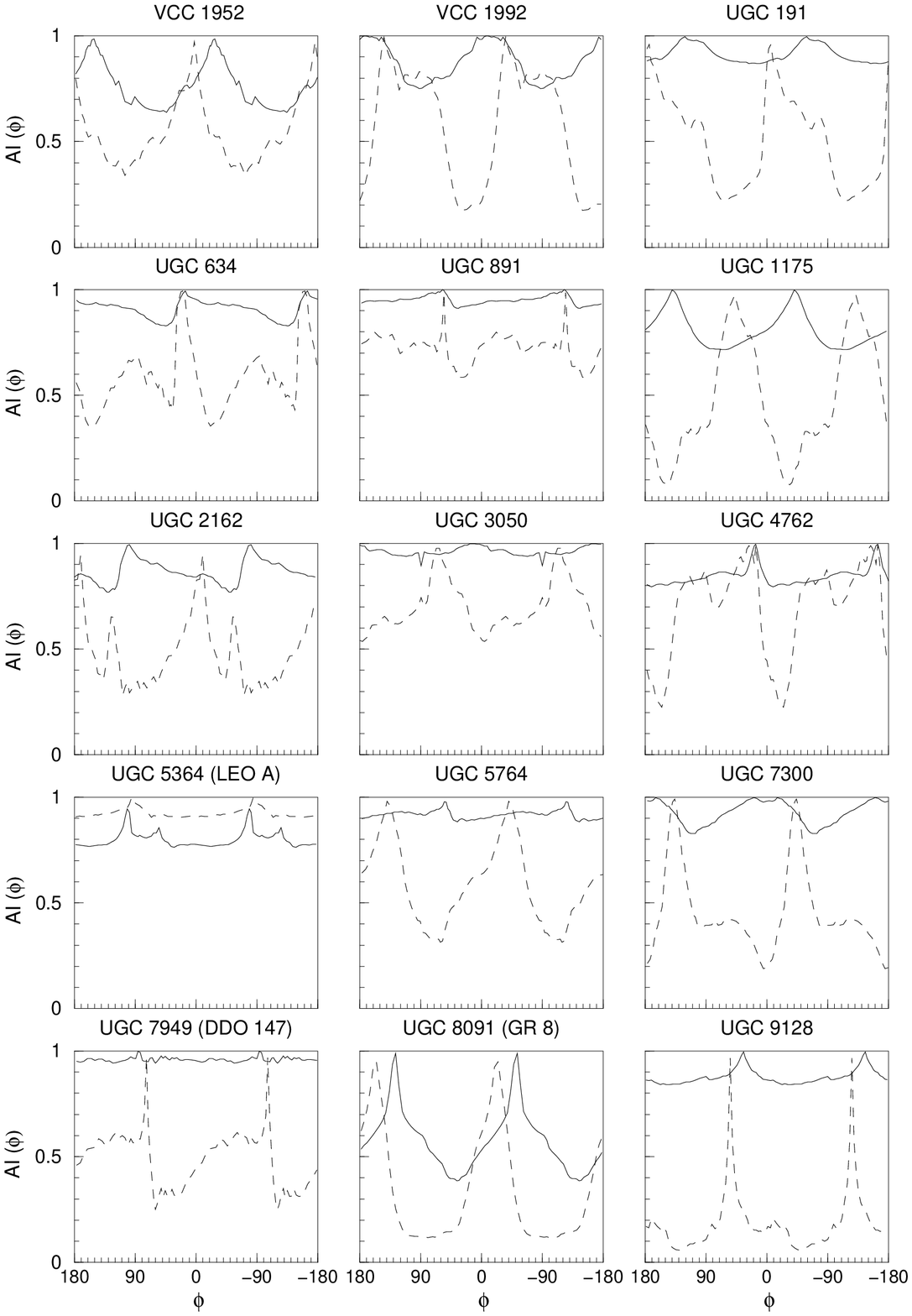}
\figcaption{Same as Fig. ~\ref{S4}. \label{S5}}
\end{figure}

\begin{figure}{}
\figurenum{2.3}
\epsscale{0.90}  
\plotone{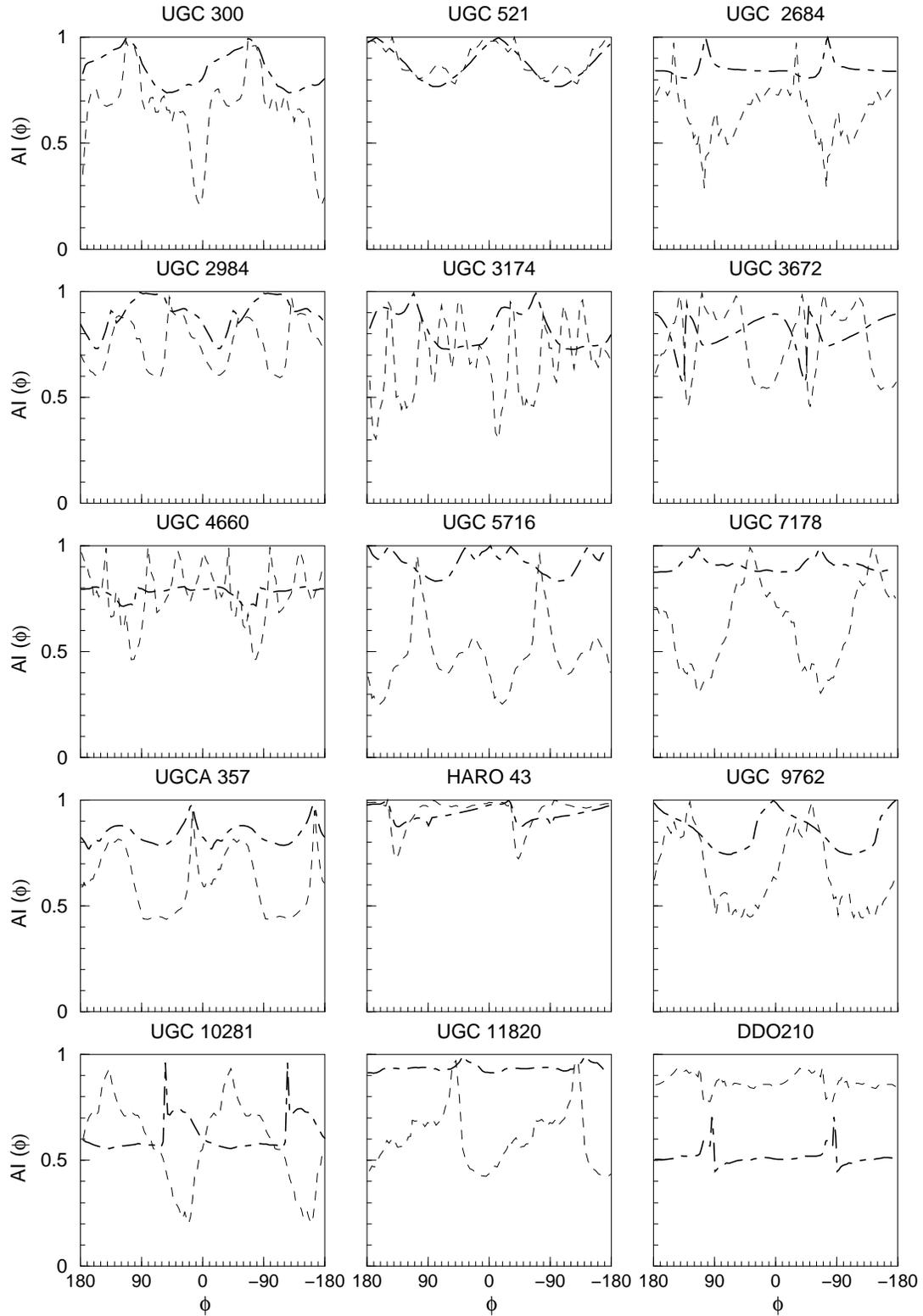}
\figcaption{Variation of AI vs. azimuthal angle
                   ($\Phi$)   for LSBs with
          B--band images for the continuum. 
          B--band is  plotted with bold dot-dashed line and H$\alpha$ 
        is plotted with a dashed line. \label{S6}}
\end{figure}

 We integrated the fluxes in  the two halves of the galaxy separated 
  by  a bisector line,  represented by the major-axis of
 the outer ellipse,  and  computed the ratio of the lower flux
to the higher flux  from the  two galaxy halves.  The resulting ratio  
  defines one asymmetry index (AI$_i$).
 After this, using the maximum allowed  90 vertexes of the ellipse contour
 produced from  the ELLIPSE task of IRAF in the plane of the galaxy, 
 we  rotated  the bisector anti-clockwise around 
 the center of the ellipse to the line defined by the next two
  opposite vertexes, and thus obtained 90  asymmetry indices,
  one for each position angle $\Phi_i$
 of the vertex.  For the maximum ellipticity (e=0.75)
   measured in the sample,
  we reach an upper spatial resolution  of $\Phi_i$= 0.4 degrees, 
  and a  lower  resolution of  $\Phi_i$=4.4 degrees. 
  For example, in a perfect circle  (ellipticity e=0)
 the resolution is   $\Phi_i$ =  4 degrees. 
This  method is more useful  than the usual one constructed
 from  two asymmetry indices because it covers 
 the full range of possibilities in azimuthal angle. Moreover,
 the presence
of  faint HII  regions  is emphasized
 by the  irregularities in  the luminosity profiles. 
 The variation of AI with position angle (the "lopsidedness
distribution", LD) is plotted for each galaxy in the sample
in Figures~\ref{S1} to  \ref{S6}.

 A representative  lopsidedness index (A) for each galaxy was  
computed by normalizing the total  lopsidedness range (the difference
 between maximum and 
 minimum AI)   to the maximum asymmetry index:
 $ A=\frac{AI_{max} -  AI_{min}} {AI_{max}}$.
 The mean  asymmetry index  $<$AI$>$,  the lopsidedness index (A),  
and  the  asymmetry amplitude  
  $ampl =\frac{ AI_{max}-AI_{min}}{2}$  are listed in Tables~\ref{T2a},
 \ref{T2b},  and  \ref{T2c}. 
A  symmetric distribution of the light is represented by
A=0 and  AI$_{max}$=1, while
   an extremely asymmetric distribution will
have A=1 and  AI$_{min}$=0.

However, galaxies in general  may  have bulge and
 disk components and, therefore,  a large range 
 in scale lengths. In order
 to enhance  the light distribution analysis
 we utilize a second parameter: the concentration index (CI).
 We  calculated the CI index 
  as the ratio of the  flux  from the inner part
 of the galaxy to   $1/3$ 
 of the flux from its outer annulus. The one-third factor
 brings the comparison to an equal-area basis, and 
 makes it  independent of the
 distance of the galaxy. The outer aperture was defined  as 
 the ellipse fitted to calculate the LD.
 The inner aperture was chosen as a smaller ellipse,
 half  the size of the outer  one. 
 The annulus is the space between the inner and the 
outer apertures.
 As defined,  CI  can
 range between zero and infinity.   

The two structural indices are similar to those used in 
 Brosch {\it et al.} (1998), with the differences being: 
(a)~the use of the $H\alpha$ flux instead of  the number counts of HII regions, 
(b)~the application of  an objective automatic algorithm instead
of eyeball recognition, 
(c)~the derivation of the full LD instead of only indices, and (d)~the use of a
normalization factor of  $1/3$ instead of  $1/4$  for CI.  Another difference is 
that here we calculate
 the CI and AI indices for the continuum, as well as for the net line emission.

\subsection{Results}

In Fig.~\ref{real2x2} we plot  the structural indices vs. the 
number of HII regions and ellipticity of the ellipse contour.
The  number of HII   regions in BCDs  ranges from one to
 three, and for LSBs from one to twelve.
 These numbers represent the number of  resolved peaks
 detected in  the LDs by an automatic algorithm that 
searches for slope changes in the LDs. 
The main limitation of  the algorithm is the
lack of resolution in special cases of
   multiple HII regions perfectly aligned in the radial direction. 
 Since the `clumpiness' of a galaxy depends on the seeing, 
the resolution at which the image  is sampled, and  on 
 the resolution  of the LDs, we cannot derive
 the number of HII regions  in real galaxies as an absolute parameter;
for example,  nearby systems will appear clumpier
 than more distant ones. 
We  show below  that a change in  the resolution, or a  difference
  in the number of HII regions between  BCDs and LSBs,
 cannot explain the differences in  concentration indices
 between   the types.
Due  to the  intrinsic irregular shape of these galaxies,
the ellipticity (e=1 - b/a) is also uncertain,
but it does provide  some   measure of the inclination. We can
 see  in   Fig.~\ref{real2x2}b and Fig.~\ref{real2x2}d that 
 the derived quantities are not simply 
the result of projection effects
 or affected by extinction through the disk.

 The profile of the lopsidedness distribution appears to be related to the
  central surface  brightness of a galaxy.
  A characteristic feature of the low surface brightness (LSB) sub-group is 
the   multi-component structure of the LD,
 with sharp features shown in  the $H\alpha$ profiles,
while the  continuum LD
is  smoother and with shallower features (Figs. \ref{S4} through \ref{S6}),
 but not  fully symmetric.  The multiplicity of  the  AI$_{H\alpha}$ profiles indicates  
that a number of individual  HII regions with different luminosities 
and sizes  are  distributed  over the galaxy;
some may  even  not be resolved  or recognized in our images but their 
contribution to the local
H$\alpha$ flux is counted by the algorithm.

 The interpretation of the LD profile widths depends not only  on the 
 sizes of the HII regions
 but also on their radial location;  single HII regions  
closer to the center  produce  a wide peak, while  those
 further out   show narrow peaks. At the same radial distance,
the bigger the size of the HII region,
the wider the LD profile will appear. A  nuclear HII region of some extent will present
a flat profile. In the  BCD sub-sample, many of
  the LDs  show profiles that are mostly  smooth,
 free of  multi component structure,
generally symmetric and wide 
  (Figs. ~\ref{S1} through  \ref{S3}), as   expected   for
  single HII regions  located near  the centers.
   BCDs tend to be more  concentrated than the LSBs;  the  median CI$_{H\alpha}$
   is 8.56 for BCDs  and  2.25 for  the  LSBs. 
 We found a strong  correlation between
  log(CI$_{Cont}$) and log(CI$_{H\alpha}$)  
  (Fig.~\ref{sym7}a) with the 
correlation coefficient cc=0.61 (F=36)\footnote{F is 
the ratio between the mean square deviation due to the regression and
the mean square deviation due to the residual variation. For a linear 
regression, which is the present situation,
F=t$^2$ and this is the equivalent of a t-test. For more details
see Draper \& Smith~(1981).}.
Linear regression tests between other data sets are listed in Table~\ref{T5}.
 At the same CI, both sub-samples  reach similar degrees of asymmetry.
Note that both galaxy  types tend to clump at 
log(CI$_{Cont}$)=0.5$\pm$0.2
and  A$_{Cont}$=0.2$\pm$0.1 (Fig.~\ref{sym7}c). 
The entire sample has a median  A$_{H\alpha}$
 of  0.69;  the median for BCDs is  0.71  and for  LSBs  is  0.69.

We find an apparent 
upper limit  for   the asymmetry of the continuum light;  97$\%$ of the
 galaxies  have A$_{Cont}  \leq $ 0.5 (Fig.~\ref{sym7}c). We also find an apparent
 lower limit  of the emission line asymmetry;   97$\%$  of the galaxies have
 A$_{H\alpha} \geq$ 0.3 (Fig.~\ref{sym7}d).  A perusal of Tables \ref{T2a}, \ref{T2b}, 
 and \ref{T2c}  shows that the
$Cont$ asymmetry is always smaller than the  H$\alpha$ asymmetry.  
A $\chi ^{2}$-test
 of the cumulative histograms
of A$_{H\alpha}$ and A$_{Cont}$ indicates that the two data sets
  originate from different distributions ($\chi ^{2}$=244,  with 18 degrees of freedom).
 The median $Cont$ asymmetry is 0.25  for BCDs, 0.21   
 for  LSBs,  and 0.23  for all the objects with narrow-band images for the $Cont$. 
Note that  objects with blue images for the continuum were not included in this analysis.   
 Median values of the structure parameters are listed in Table~\ref{T6}.

 A fundamental issue is whether the continuum
and line-emission LDs are correlated  in angular phase.  We should expect
a correlation if the locus of recent star formation, as witnessed
by the  H$\alpha$  emission, responds  with a delay to some 
 disturbance of the stellar distribution
  (the continuum light), producing a lag in the angular distribution of the
azimuthal indices. This can be understood  in a scenario of
 rotating disk-like systems. In fact,  HI synthesis maps of a number these
galaxies (Skillman~{\it et al.}~1987; van Zee~{\it et al.}~1997c,
1998a,~b) 
 show rotation-dominated systems with maximal rotation velocities of
 40 -- 100 km sec$^{-1}$    
  and  with    slowly rising  rotation curves,
typical  of  very late-type  spirals (some appear to be undergoing 
small differential rotation),
or systems with velocity still increasing
  beyond the optical disk, characteristic of the  solid-body rotation
  found in many low-mass systems. 
  In those cases, the angular phase correlation may depend on many factors,
   such as
  differences in  the angular momentum of the stellar and  gas masses,
  rotational  speed,   disk shear,  and external SF triggers. 
 Note that  
 GR~8, Leo~A and DDO~210 do not have well defined rotation
curves (Carignan~{\it et al.}~1990; Young \& Lo~1996;
 Young, van Zee, \& Lo~2000).
The mismatch of the
  velocity gradient and the HI major axis in Leo A hints that 
 some very low mass systems may be tumbling 
rather than spinning. In such cases,   a delay between the past and 
 the recent onset of star formation should also be expected. 

 To measure the correlation in the phase space
 of the line azimuthal asymmetry
 distribution with the  distribution of the continuum light 
we used  a similar analysis  to that applied
to the study of AGNs variability in the time-frequency
domain (Netzer~{\it et al.}~1996;  Kaspi~{\it et al.}~2000). 
The technique is the derivation of the
 cross-correlation~function~(CCF), which is a set
of correlation coefficients, giving a measure of the correlation
between two data sets.  We used here two methods:
the first one is the discrete~correlation~function~ 
(DCF;~Edelson \& Krolik~1988), which we applied
 after interpolating 45 continous data points
 every $4^{\circ}$. 
The second method is
 the Z-transformed discrete correlation function
(ZDCF) of Alexander~(1997), which is an improvement
of the DCF. For unevenly-sampled
sets of data  the ZDCF has the advantage  
that it avoids interpolation  and reduces
the resulting uncertaintly in the position of the peak. This
is  a consequence of the  Fisher~Z-transformation to the correlation
 coefficients and of the binning   by  equal   population,
 rather than by equal separation.

 The typical errors
in the lags rage between $0.5-2 \arcdeg$, with the exception of
 $40 \arcdeg$   in
IIZW40. The two methods (DCF and ZDCF)
gave consistent results for our data,
and we will refer to the ZDCF results in the following analyses.
The uncertainties in the cross-correlation lags were 
conservatively over-estimated by the  Monte Carlo-averaged
ZDCF with simulated random errors and were
 provided by the ZDCF procedure  of Alexander~(1997).
The results of the cross-correlation analysis are presented
in Tables~\ref{T2a} and \ref{T2b}.
The columns labeled ``r$_{zdcf}$'' show the   
peaks  of the CCFs,   defined as the point of maximum
correlation; 
a high value of  ``r$_{zdcf}$'' implies a good correlation 
between the two azimuthal indices at the listed  ``lag''.
For the definition of ``r$_{zdcf}$'' see  Alexander~(1997).
The sign of the lag is defined as AI$_{Cont}$ - AI$_{H\alpha}$,
that is  AI$_{Cont}$ lags after  AI$_{H\alpha}$.

The cross-correlation (CC) analysis  of the H$\alpha$  vs. 
continuum distribution of AIs  
 indicates a very high CC   for a broad range of
angular phase lags.  The CC is higher for BCDs
than for LSBs, and there is  a trend for  smaller angular phase
 lags in  BCDs than in  LSBs  (Figs.~\ref{cross}a and
  \ref{cross}b). In fact, $\sim 62 \%$  of the BCDs
 having peak CC coefficient above 0.8  show
  lags smaller than $|\Delta\Phi|  < 30 \arcdeg$
compared to  $\sim 33 \%$ of LSBs.
 The higher CC  is  explained  by increased
CIs  (Fig.~\ref{cross}c), however 
the distribution of   lags seems to be
independent of the concentration parameter
 (Fig.~\ref{cross}d). 

 \section{A  random distribution of star formation regions?}
In this section we explore the possibility that the properties found
for star-forming regions in dIs can be produced by random processes
that engulfs the full scale of a   galaxy.
 We tested a  model  of random star formation  by constructing 
   1000 images of  galaxies,  which simulate the observed 
 net H$\alpha$-flux and off-band  red emission of   dIs as found above, without
  distinguishing between LSB and BCD types.
 The model was created with the ARTDATA
 package in IRAF and included 
      atmospheric  seeing effects and detector readout noise.

 A  galaxy  was modeled as a disk  centered on a  256$\times$256 pixel image
 with zero background. The intensity profile was that of an exponential disk
$I = I_\circ exp(-1.6783 R/R_\circ)$, with the scale radius R$_{\circ}$
containing  half  the total flux. 
The apparent integrated magnitudes, scale-lengths,  position angles of the major axis,
 and ellipticities were allowed to change randomly. 
 The  total magnitudes followed a Schechter (1976) luminosity function
with $\alpha$=1.6 and
    $M_\star=-21.41$ in the red continuum,  
 covering the apparent magnitude range from 17 to 19,
 similar to that of  the objects in our sample.  The maximum  semi-major
axis at half-flux  was set to
30 pixels. 
   The ellipticity was allowed to  vary     between 
 0.05  and  1.00. Random noise was added to the image by using Poisson statistics;
a similar process was followed for the net and continuum images described below.
  At this stage, the output   parameters of  the disk were recorded as
an ellipse contour with a semi-major axis twice the derived scale length. That is,
CI$_{Cont}=3$ and A$_{Cont}=0$,  by the definition of the underlying exponential disk.

The  $H\alpha$  emission image was created by random generation 
 of coordinates    of  up to 15 objects  within the  ellipse
 derived from the disk on a mean zero background.
This range (1-15) covers the
  number of  resolved peaks detected in  the LDs with
 the algorithm, as
explained before. We will show 
 that changing the total resolution, or  the ratio between maximum
and minimum resolution, cannot explain
 the differences in  concentration indices between BCDs and LSBs.
 The objects simulate HII regions, whose apparent
 magnitudes  were allowed to  change randomly between 18 and 23
 following   a  shallow  power law with index 0.1. This range of
 magnitudes reproduces the $H\alpha$   flux densities  observed
 for the   HII regions of our sample of galaxies   
and  yields total line fluxes 
 in the range 10$^{-15}$ - 10$^{-13}$ erg cm$^{-2}$ s$^{-1}$. 
We assumed a simplified profile
  for an individual HII cloud as  a spherical  distribution
with a star-like Moffat  profile ($\beta$ =2.5).  A Moffat profile 
(Moffat 1969) appears  
more natural than a  Gaussian,  because it produces  a sharper boundary 
to an HII region,
as expected for a Str\"{o}mgren sphere. However, this choice does not 
 appear to affect the results described below.
 
The red continuum image  was simulated by adding to the
   smooth underlying exponential disk the same  list of objects
 coordinates  and  flux densities that represented 
 the HII regions, but this time simulating   red star clusters (or super
 star clusters) distributed on the disk.
Keeping the same
distribution (with zero  angular phase lag) implies an  a-priori  correlation of
 HII regions with the red star clusters restricted to zero  angular phase lag.
Keeping the same flux densities  implies  a
uniform  equivalent width (EW) for all individual HII regions,  limited to the
 FWHM of the narrow filters used for the observed galaxies, that is 50 - 89\AA\,. This
assumption is justified by our finding for dIs
in the Virgo Cluster  where we showed that  individuals HII regions
are restricted to EW=10 to 100\AA\   (Heller {\it et al.}~1998).
  We found that, in this way,  the
 images and the LD profiles of the simulated  net and  continuum images   
reproduced the patterns observed in  the real images.
The simulation results in high degrees 
of  star formation lopsidedness with  a median A$_{H\alpha}=0.77$, 
  for CI$_{H\alpha}$ ranging from 0.01 to 30  (median CI$_{H\alpha}=1.05$).  
  A comparison set of net H$\alpha$ 
 and  continuum images, as well as plots of the azimuthal asymmetry  of
a real galaxy and a simulated one is shown in Fig.~\ref{example}.

We plot in Fig.~\ref{model2x2} the dependence of  A$_{H\alpha}$  and   CI$_{H\alpha}$ 
of the simulated galaxies on the
number of HII regions (N)  and on the integrated  H$\alpha$ fluxes. 
The results show
no dependence on the  total flux. Changing  N does not affect   the lopsidedness  
range of possibilities, but  there is a clear  trend to  CI$_{H\alpha}$=1    as the number
of HII regions increases. This effect  is reflected in Figure ~\ref{modelall} 
  where we plot
    A$_{H\alpha}$  vs. log(CI$_{H\alpha}$)   for all simulated galaxies
 (filled circles). Note that
these galaxies are distributed around  CI$_{H\alpha}$=1.  The actual galaxies 
(represented by triangles and squares) show, in general,  higher 
 values of CI$_{H\alpha}$  than the simulated galaxies.

  A closer look at  the  plots of    A$_{H\alpha}$  vs. log(CI$_{H\alpha}$)
for different number of HII regions (Fig.~\ref{model12}) helps
  us interpret this effect. 
We see that the ``phase space'' accessible to simulated galaxies in the
 AI-CI plane becomes more restricted, the more HII regions a galaxy has.
While for N=1 objects almost one half of the plane is populated,
at N=12 the distribution is concentrated mostly at CI$_{H\alpha}$=1 for a large spread of AI's.

 A trend of reduced concentration with increasing number of
HII regions  is visible for the right side of   the distribution in Fig.~\ref{model12}.
 This is explained as the result of the fact that
 the more HII regions a (simulated) galaxy has, the more 
``balanced'' is the distribution of these HII regions.
Another trend is visible for the left side of the distribution   in Fig.~\ref{model12}.
The fewer HII regions a galaxy has, the better the chance to find these
``unbalanced'', more to one side of a galaxy than the other. This means that
galaxies with few HII regions will  be more asymmetric than
galaxies with many HII regions.
A test for 15-20  HII regions did not change
the distribution for N=12, but it is obvious that by increasing N  
 it will finally converge  to the point  (A$_{H\alpha}$=0,
 CI$_{H\alpha}$=1).

In order to test if the number of HII regions of the actual galaxies
was exaggerated by the number of irregularities detected in the LDs
 we plotted in  Fig.~\ref{model123} only simulated galaxies 
with N=1, 2 and 3.  We can see that
reducing the number of HII regions shifts the simulated galaxies to
   a  higher mean  CI$_{H\alpha}$=1.3. 
This fits  better  the LSB sub-group,  however,
 it is not enough in     order to explain the general shift of the BCD galaxies. 
We discuss this  in the next section.

Summarizing, we  have shown  how the degree of asymmetry  
and concentration index
of star forming regions in simulated galaxies  change with 
 the total number of HII regions and their luminosity distribution.
 A similar asymmetry  behaviour occurs for the continuum,  
but to a lesser degree, due to 
the relatively smaller contribution of the young stellar clusters
  over  the disk brightness.

 \section{Discussion}
  Our analysis indicates that most  dIs show a  lopsided
 morphology in their recent  star formation and in the
 distribution  of red light. Since the analysis was performed
on the basis of structural indices that are  independent of 
  distance,  angular size,  and/or   inclination 
of the galaxies, we believe that this is a intrinsic property
of dwarf-irregular galaxies. 
The entire sample has a median lopsidedness index of 0.69
  in their star forming distribution;  similar results are obtained
 for LSBs and BCDs. For the same concentration index,
 LSB and BCD galaxies   reach  similar    degrees of lopsidedness.
 The correlation detected   between  the continuum
  and the line emission concentration
 is supported by a strong correlation
 between  on-going star formation 
 regions and    the red stellar population in  the angular-phase domain. 
The correlation is stronger in BCDs, with  a trend
 for  smaller angular phase lags  than  LSBs.
The results are consistent with the  correlation found between  line
 and continuum fluxes 
of individual HII regions in dIs in the Virgo Cluster, which is
 much stronger for BCDs than for LSBs~(Heller {\it et al.}~1999).

We   mentioned already an important difference
 between the  BCD and LSB galaxies: 
BCDs    exhibit  stronger  H$\alpha$
 concentration  in their nuclear  regions than do  LSBs. 
 This  is  emphasized by the profiles of the lopsidedness
 distribution and is best seen in  the distribution
of  CI$_{H\alpha}$ (see e.g., Fig.~\ref{sym7}a), 
 where the squares tend  more to the right than the 
 filled triangles. This tendency is not   shown  as strongly in the
distribution of  CI$_{Cont}$; the degree of  concentration
 of the red continuum light in BCDs is  rather similar to 
that in LSB galaxies (Fig.~\ref{sym7}c).

Interpreting the continuum light in the simulated galaxies as 
showing the distribution of previous stellar generations 
implies that in this aspect LSBs and BCDs are similar,
but it is clear  that  shorter scale-length  exponential disks are
  needed  in  compact   galaxies.  The implication is the
 presence of at least one past major star formation event in the
 central regions.  This result is consistent with detailed
surface brightness fitting of BCDs (e.g., Salzer \& Norton 1998, 
Norton \& Salzer 2000). 

The differences  become more evident as one compares the distribution
 of  newly-formed stars, as measured by the H$\alpha$ emission
 in the real galaxies,  with that in the simulated galaxies.
 BCDs have more concentrated emission and do not fit 
the median CI$_{H\alpha}$ 
of the simulated galaxies. 
 The different concentration indices
measured for the two types are not  merely a result of 
fewer HII regions per BCD than per LSB; reducing to 1-3 the
 number of HII regions in simulated galaxies
  increased  the mean of the distribution by 30\%, 
but     the lower limit in CI$_{H\alpha}$ for BCDs stayed  were
 the model predicts the mean.   We conclude that    randomly
generated  star formation  may be
proceeding through the disk in LSB dwarf-irregular galaxies, 
 but  probably not in BCDs.  
 
The  higher correlation of the line and continuum LDs
in BCDs is reminiscent of
 the similarity of blue and near-IR  images of galaxies in the HDF
 (Richard Ellis, private communication).
In the rest frame of these galaxies, imaged with the 
WFPC-2 and NICMOS, the optical colors correspond to
 rest-frame UV (i.e., the young stars) and the
near-IR correspond to the optical continuum light used
 here to estimate the distribution of the older stellar populations. 
Their correlation shows that at z $\approx$2 the young
stars form where there are more of the old stars, as we found here
for dwarf irregulars.

From the  diagram  in Fig.~\ref{model123} we  learn that
 there is a limit to the degree of concentration a  real  LSB galaxy
 can have; higher CI indices would imply
non-realistic, extremely extended galaxies with a 
dominant optical core, such as the very rare Malin I-types. 
In our LSB sample,
the upper  concentration limit    is set  by  the  extended-HI
 galaxy sub-group of van Zee~{\it~et~al.}~(primary). 
The behavior should be similar at the lower limit, but this
 limit is not well-defined.

 \section{Conclusions}
We analyzed  images of  78 dIs and
measured the concentration and asymmetry of the  H$\alpha$
line and  red continuum emission by applying  an objective automatic
algorithm and by tracing the asymmetry along the azimuthal direction. 
Our findings show a   high degree of asymmetry of the H$\alpha$
emission,  which follows  a milder asymmetry in the distribution of the
 red  light. Both concentration  indices (line and continuum) 
are highly  correlated. 
The continuum and line emission lopsidedness distribution
are correlated in angular phase, and there is  a trend for
higher correlation and smaller angular 
phase lags in BCDs than in LSBs. 
  We found considerable differences
between  these two types of dwarf galaxies
 in terms of lopsidedness distribution profiles and
 concentration index.

We showed that a random distribution of HII regions
 can produce the observed lopsidedness of low surface brightness
 disk-like systems. 
The key parameters that most affect  the model are the scale
lengths,  the number of  HII regions per galaxy, and 
the luminosity distribution  of the HII regions.
The model  fits  well most of the observables:
 the  frequency, strength,  and  profiles of the lopsidedness 
  are  recovered, both in the line emission and in the continuum.
 The model  matches the distribution observed
in normal LSB dwarf galaxies in the lopsidedness-concentration plane,
 but  short scale-length exponential disks and  some 
central diffuse light components are called   for in  
LSB galaxies with extended HI envelopes.
We  showed that reducing the number of HII regions cannot
explain the higher concentrations observed in BCDs.
 It seems that a random  distribution of star formation may 
explain the patterns  observed in  LSB dIs, but not in BCDs.
\section*{Acknowledgments}
AH and NB acknowledge 
support from the
US-Israel Binational Science Foundation. EA is supported by a special grant from 
the Ministry of Science and the 
Arts to develop TAUVEX, a UV space imaging experiment.  NB is grateful for the continued 
support of the Austrian Friends of Tel Aviv University.  Astronomical 
research at Tel Aviv University
is partly supported by a Center of Excellence Award from the Israel Academy of 
Sciences.

\section*{References}

\begin{description}

\item Abraham, R. G., van den Bergh, S., Glazebrook, K.,  Ellis, R. S.,
 Santiago, B. X., Surma, P., \& Griffiths, R. E., 1996, ApJS, 107, 1.

\item Alexander, T.,  in ``Astronomical Time Series'', edited by D. Maoz,
 A. Sternberg \& E. M Leibowitz, 1997, Astrophysics and Space Science Library, 218,
Kluwer: Dordrecht.

\item Almoznino, E. \& Brosch, N., 1998, MNRAS, 298, 920.

\item Brosch, N., Heller, A.B. \& Almoznino, E., 1998, MNRAS, 300,1091.

\item Carignan, C., Beaulieu, S., \& Freeman, K. C., 1990, AJ, 99, 178.

\item Conselice, C. J., 1997, PASP, 109, 1251.
\item Conselice, C. J.  \& Bershady, M.A., in ``After the Dark Ages: 
    When Galaxies were Young (the Universe at $2 < z <5$).", 9th Annual
 October Astrophysics Conference in Maryland, held 12-14 October,1998. ,
edited by S. Holt \& E.  Smith.,
 American Institute of Physics Press, 1999,  225.

\item  Dohm-Palmer, R. C., Skillman, E. D., Saha, A., Tolstoy, E., Mateo, M.,
       Gallagher, J., Hoessel., J.,  Chiosi, C., Dufour, R. J., 1997, AJ, 114, 2527.
\item  Dohm-Palmer, R. C., Skillman, E. D., Saha, A., Tolstoy, E., Mateo, M.,
       Gallagher, J., Hoessel., J.,  Chiosi, C., Dufour, R. J., 1998a, AJ, 115, 152.
\item  Dohm-Palmer, R. C., Skillman, Gallagher, J., Tolstoy, E., Mateo, M.,
       Dufour, R. J., Saha, A., Hoessel., J., Chiosi,  C., 1998b, AJ, 116, 1227.

\item Draper, N.R. \& Smith, H., 1981, {\it Applied Regression
Analysis}, New York: Wiley,  32.

\item Edelson, R., \& Krolik, J. H., 1988, ApJ, 333, 646.

\item Elmegreen, B. G., 1998, in "Origins of Galaxies, 
Star, Planets and Life" (C. E.  Woodward,
 H. A. Thronson  \& M. Shull, eds.), ASP series.

\item Gerola, H., Seiden, P. E., Schulman, L., S., 1980, ApJ, 242, 517.

\item Heller, A. B., Almoznino, E. \&  Brosch, N.   1998, IAU Colloquium No. 171: 
   ``The Low Surface Brightness Universe", ed.  J. I. Davies,  C. Impey,
    \& S. Phillips,  ASP Conf. Series, 170,  282.
\item Heller, A. B., Almoznino, E. \&  Brosch, N.,   1999, MNRAS, 304, 8.
\item Heller, A. B., Almoznino, E. \&  Brosch, N.,   2000, in preparation.

\item Hunter, D. A., Elmegreen, B. G., \& Baker, A. L., 1998, ApJ, 493, 595.

\item  Jog, C. J., 1997, ApJ, 488, 642.

\item Kaspi, S., Smith, P. S., Netzer, H., Maoz, D., Jannuzi, B. T., \& Giveon, U.,
   2000,  ApJ, in  press.

\item Kornreich, D. A., Haynes, M. P., \& Lovelace, R. V. E., 1998, AJ, 116, 2154.

\item Larson, R. B., 1986, MNRAS, 218, 409. 

\item Leitherer, C. \& Heckman T.M., 1995, ApJS.  96, 9.

\item Levine, S. E. \& Sparke, L.S., 1998, ApJ, 496L, 13L.

\item Netzer, A., Heller, A., Loinger, F., Alexander, T., Baldwin, J. A.,
   Wills, B. J., Han, M., Frueh, M., \& Higdon, J. L., 1996, MNRAS, 279, 429.

\item Mihos, J. C., McGaugh, S. S.,  \& de Blok, W. J. G., 1997, AA, 477, L79.

\item Moffat, A. F. J., 1969, A\& A, 3, 455.

\item Norton, S. A. \& Salzer, J. J., 2000,  in preparation.

\item Rudnick, G.  \& Rix. H.W., 1998, AJ, 116, 1163.

\item Salzer,  J. J. \& Norton, S.A., 1998,  IAU Colloquium No. 171: 
   ``The Low Surface Brightness Universe",  ed.  J. I. Davies,  C. Impey,
    \& S. Phillips,  ASP Conf. Series, 170,  253.

\item Schechter, 1976,  ApJ, 203, 297.

\item Skillman, E. D., Bothun, G. D., Murray, M. A.,  \& Warmels, 
 R.H., 1987, A\&A , 185, 61.

\item van Zee, L., Haynes, M. P.,  \& Salzer, J.  J.,  1997a, AJ, 114, 2479.
\item van Zee, L., Haynes, M. P.,  \& Salzer, J.  J.,  1997b, AJ, 114, 2497.
\item van Zee, L., Haynes, M. P.,  Salzer, J.  J.,  \& Broeils, A. H., 1997c, AJ, 113, 1618.
\item van Zee, L., Westpfahl, D., Haynes, M. P.,  Salzer, J.  J., 1998a, AJ, 115,1000.
\item van Zee, L., Skillman, E. D.,  \& Salzer, J.  J., 1998b, AJ, 116,1186.

\item Walker, I. R.,  Mihos, J. C. \&  Hernquist, L. , 1996, ApJ, 460, 121.

\item Young, L. M., Lo, K. Y., 1996, ApJ, 462, 203.
\item  Young, L. M., van Zee, L., \& Lo, K. Y.,  2000, in preparation.

\item Zaritsky, D.  \& Rix. H. W., 1997, ApJ, 477, 118.

\end{description}

\newpage
\begin{table}{}
\vspace{17cm} 
\includegraphics{table1ab.ps}
\includegraphics{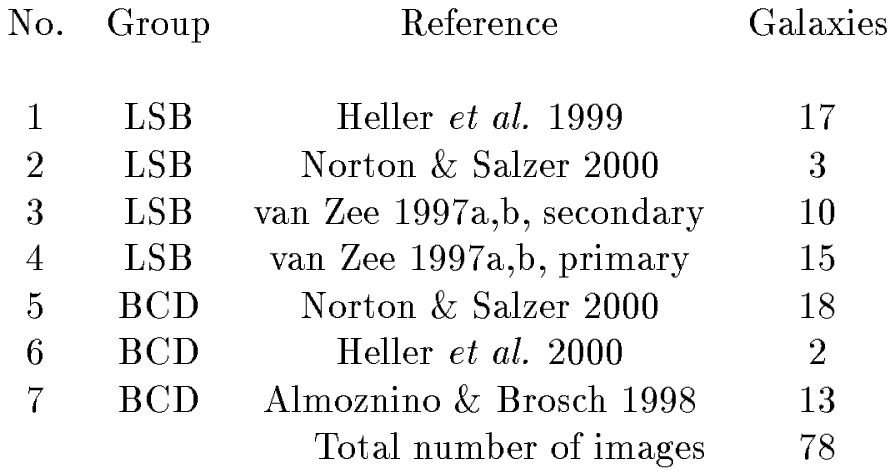}
\caption{Sample of galaxies. \label{T1abc}}
  \tablecomments {The various symbols are:\newline
             a = semi-major axis  in arc seconds.\newline
              $e=1-\frac{b}{a}$  ellipticity of aperture.\newline
              PA = position angle  in degrees, measured counter-clockwise from +y.\newline
              (*) certain/or probably interacting galaxies.\newline
             (+) objects with B--band images  for the continuum.}
\end{table}

\newpage
\begin{table}{}
\vspace{17cm}
\tablenum{2.1}  
\includegraphics{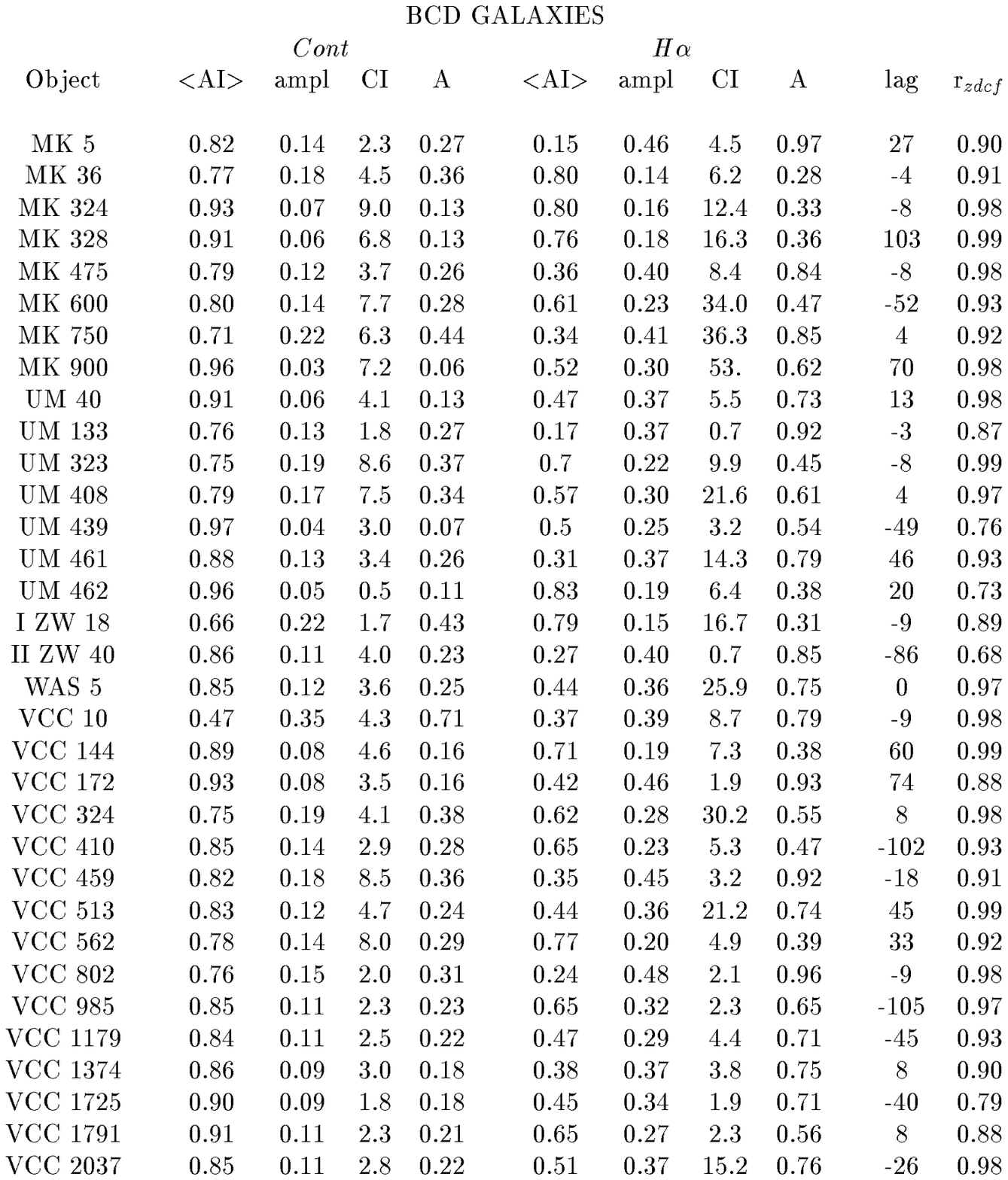}
\caption{Structural parameters for  BCD galaxies in the sample. \label{T2a}}
\tablecomments {The various symbols mean:\newline
 $<$AI$>= < \frac {flux_i} {flux_j} >$                    mean  asymmetry index. \newline 
 ampl $= \frac{ AI_{max}-AI_{min}}{2}$               asymmetry amplitude.\newline
 CI$=\frac {flux_{in}} {\frac{ flux _{out}}{3}}$    concentration index. \newline
  A$=\frac{AI_{max} -  AI_{min}} {AI_{max}} $    lopsidedness index.\newline
  r$_{zdcf}$  is the maximun cross-correlation-coefficient of  AI$_{H\alpha}(\Phi)$ with 
$AI_{Cont}(\Phi)$ for  $\Phi$=lag in~degrees.}  
\end{table}
\newpage
\begin{table}{}
\vspace{17cm}
\tablenum{2.2}  
\includegraphics{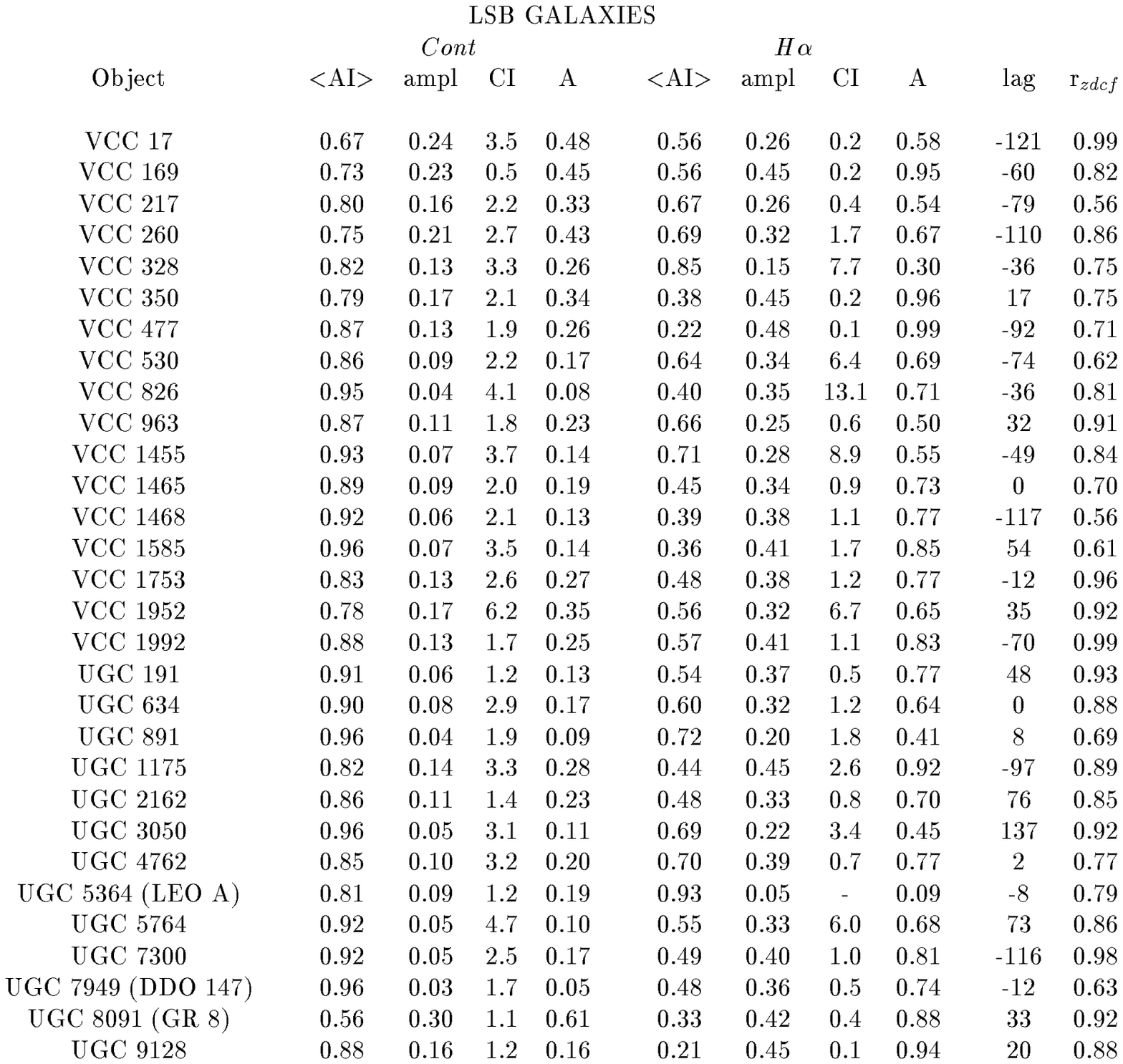}
\caption{Structural parameters for  LSB galaxies with narrow-band  $H\alpha$ 
               images for the continuum distribution. \label{T2b}}
\end{table}

\newpage
\begin{table}
\vspace{10cm} 
\tablenum{2.3} 
\includegraphics{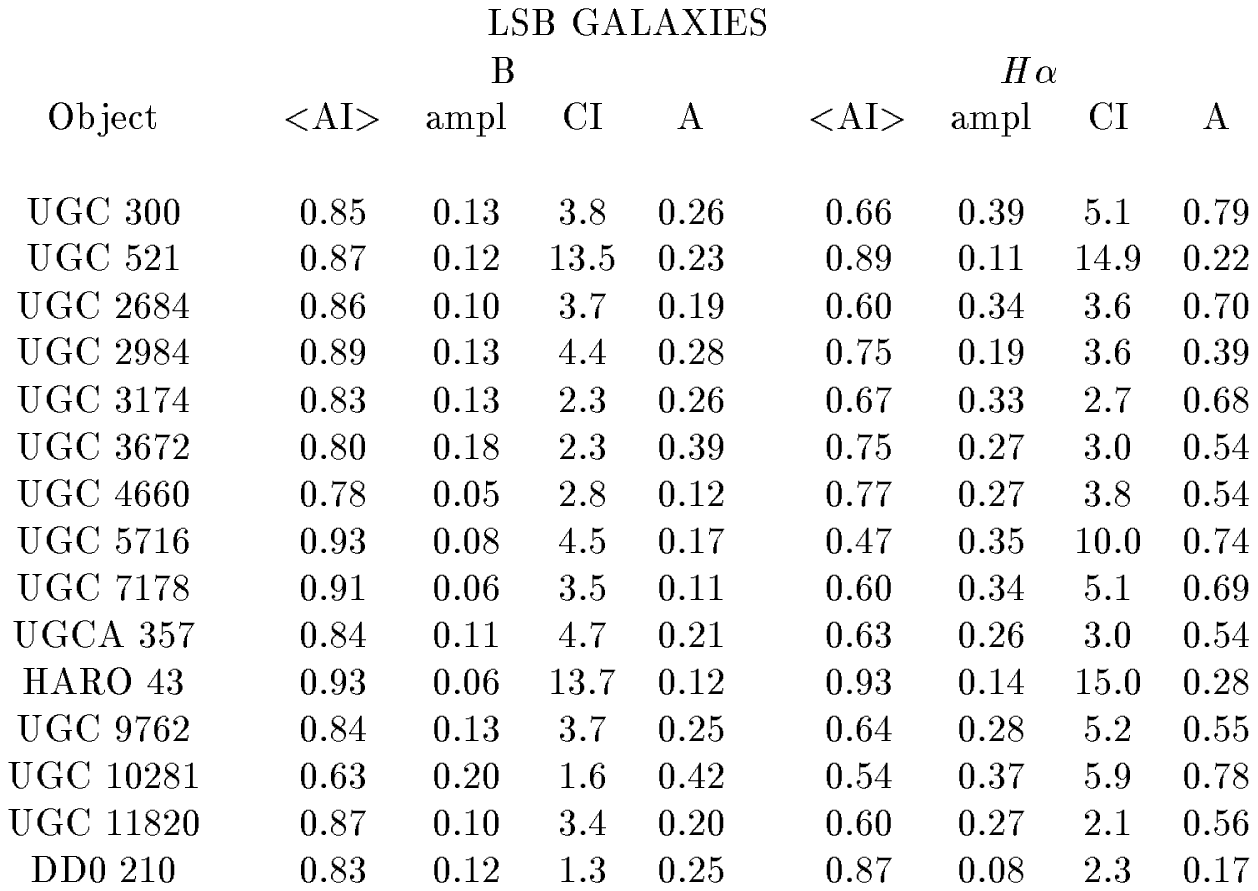}
\caption{Structural parameters for LSB galaxies with B--band  images for
                the continuum distribution. \label{T2c}} 
\end{table}
\setcounter{table}{2}
\clearpage
\newpage
\begin{table}{}
\vspace{5cm}  
\begin{tabular}[b]{cccc}
\multicolumn{1}{c}{Var. 1} & 
\multicolumn{1}{c}{Var. 2} & 
\multicolumn{1}{c}{cc } & 
\multicolumn{1}{c}{F} \\
\tableline
log(CI$_{Cont}$)         &  log(CI$_{H\alpha}$)       & 0.67 & 60\\
A$_{Cont}$               & A$_{H\alpha}$              & 0.16 & 1.9\\
A$_{H\alpha}$            & log(CI$_{H\alpha}$)        & -0.40 & 14\\
A$_{Cont}$               & log(CI$_{Cont}$)           & -0.02 & 0.03\\
\tableline
\end{tabular}
\caption{Linear regression tests. \label{T5}}
\end{table}

\begin{table}{}
\vspace{5cm}
\begin{tabular}[b]{cccc}
\multicolumn{1}{c}{Var.} & 
\multicolumn{1}{c}{BCD} & 
\multicolumn{1}{c}{LSB } & 
\multicolumn{1}{c}{BCD+LSB} \\
\tableline
CI$_{H\alpha}$ &   8.56        & 2.25         & 4.23    \\
CI$_{Cont}$    &   4.90        & 2.70         & 3.43    \\
A$_{H\alpha}$ &    0.71      & 0.69        & 0.69\\
A$_{Cont}$    &    0.25      & 0.21        & 0.23\\
\tableline                        
\end{tabular}
\caption{Median values of the structure parameters. \label{T6}}
\end{table}
\newpage

\newpage

\newpage

\newpage

\newpage

\newpage
\setcounter{figure}{2}

\newpage
\begin{figure}{}
\epsscale{0.75}
\plotone{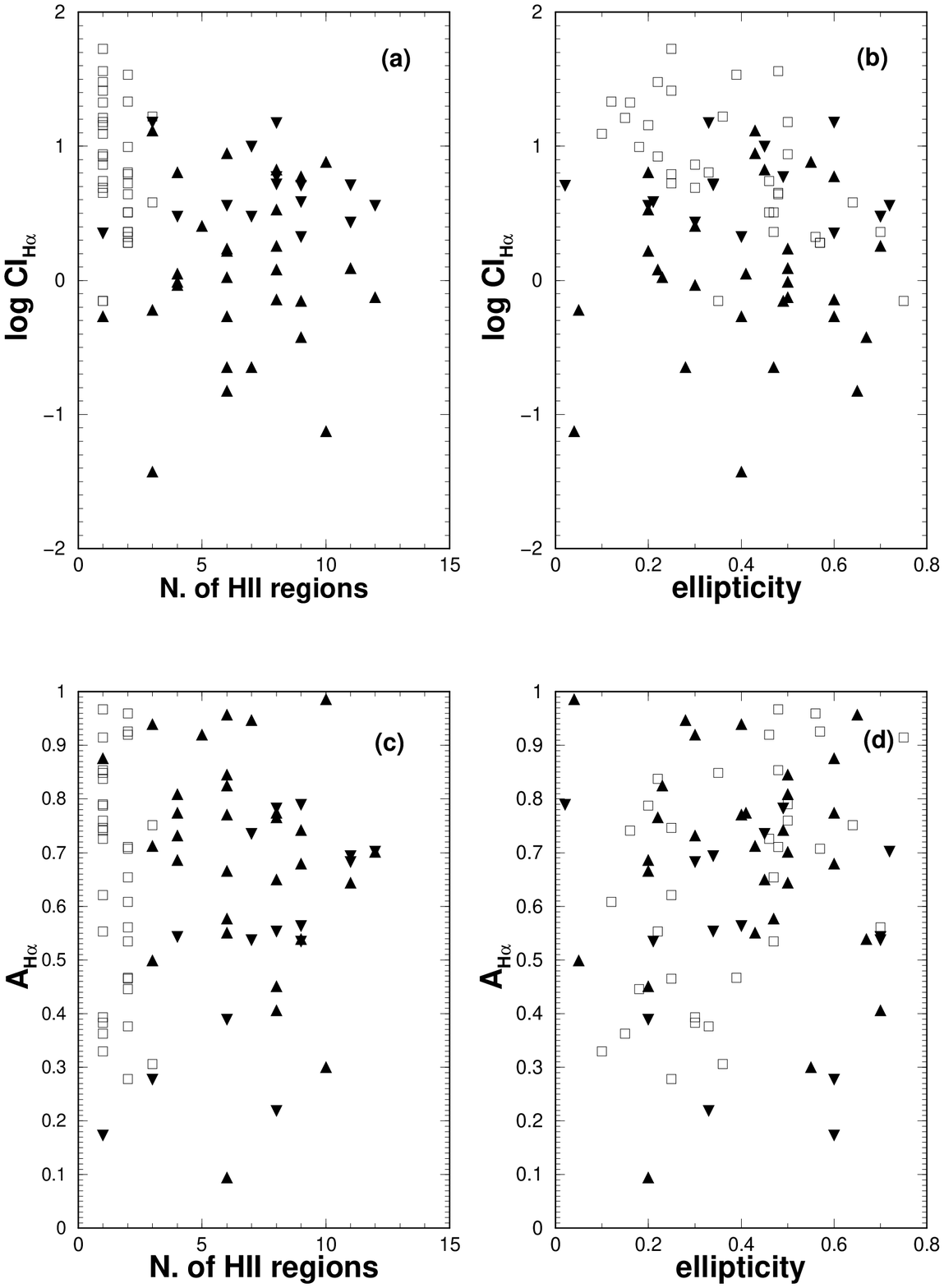}
\figcaption{Observed galaxies:  BCDs are represented by squares, 
                        LSBs are represented by triangles-up for objects with
                        narrow-band  $H\alpha$  images    for   $Cont$, and by
                      triangles-down for objects with  B-band images for $Cont$. 
                      The different panels show
                       concentration index of $H\alpha$ flux vs. 
                        number of HII regions in (a), 
                      concentration index of $H\alpha$ flux vs. 
                    ellipticity in~(b),  lopsidedness vs.  number of HII regions in (c), and 
                   lopsidedness vs. ellipticity in~(d).  \label{real2x2}}
\end{figure}

\newpage
\begin{figure}{}
\plotone{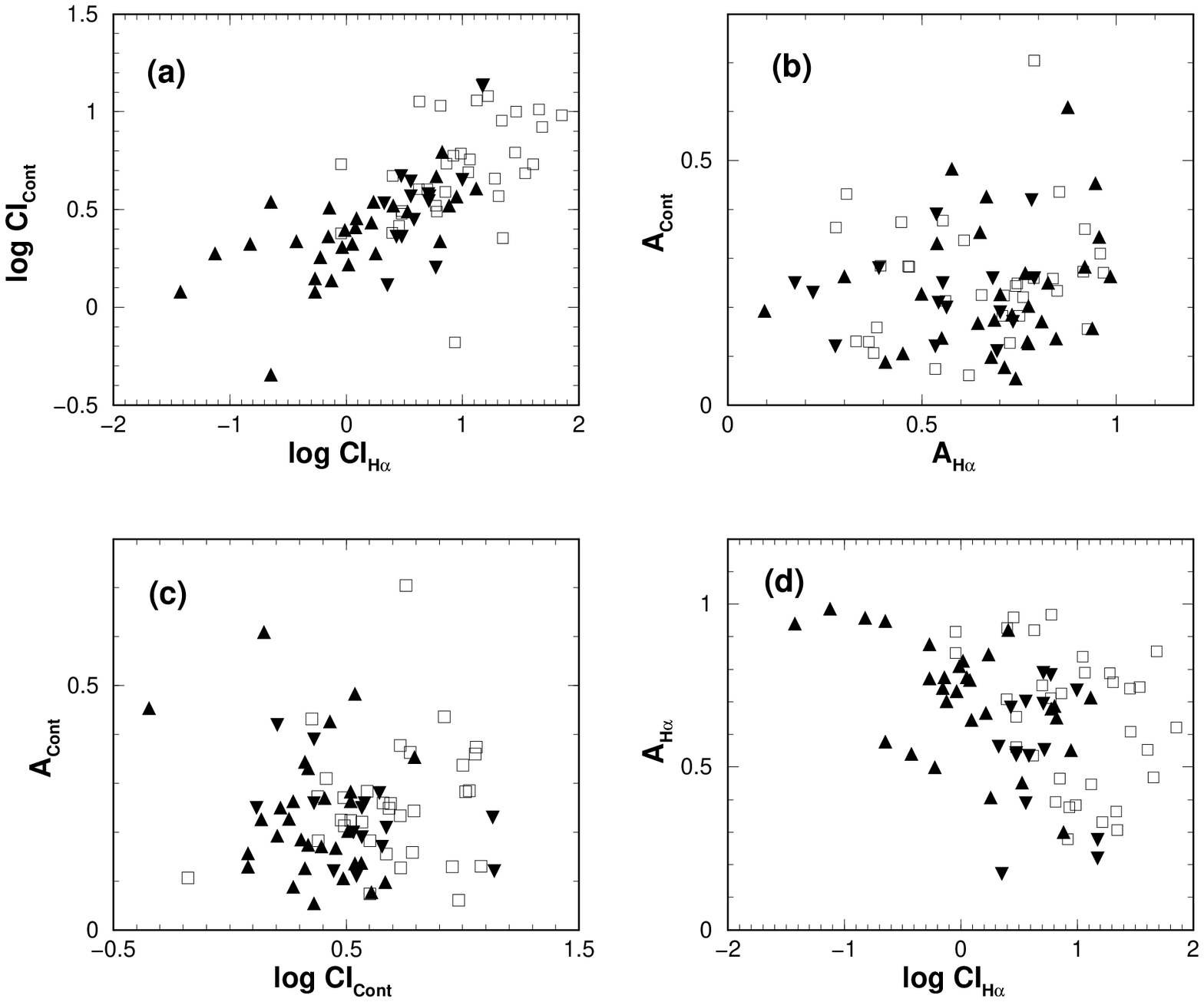}
\figcaption{Correlations of parameters: 
                     the different panels show
                    log~CI$_{Cont}$ vs. log~CI$_{H\alpha}$ in (a),
                    A$_{Cont}$ vs.  A$_{H\alpha}$ in (b),
                     A$_{Cont}$ vs. log~CI$_{Cont}$ in (c), and
                    A$_{H\alpha}$~vs. log~CI$_{H\alpha}$ in (d). 
                   Symbols as for Fig.~\ref{real2x2}. \label{sym7}}
\end{figure}

\newpage
\begin{figure}{}
\plotone{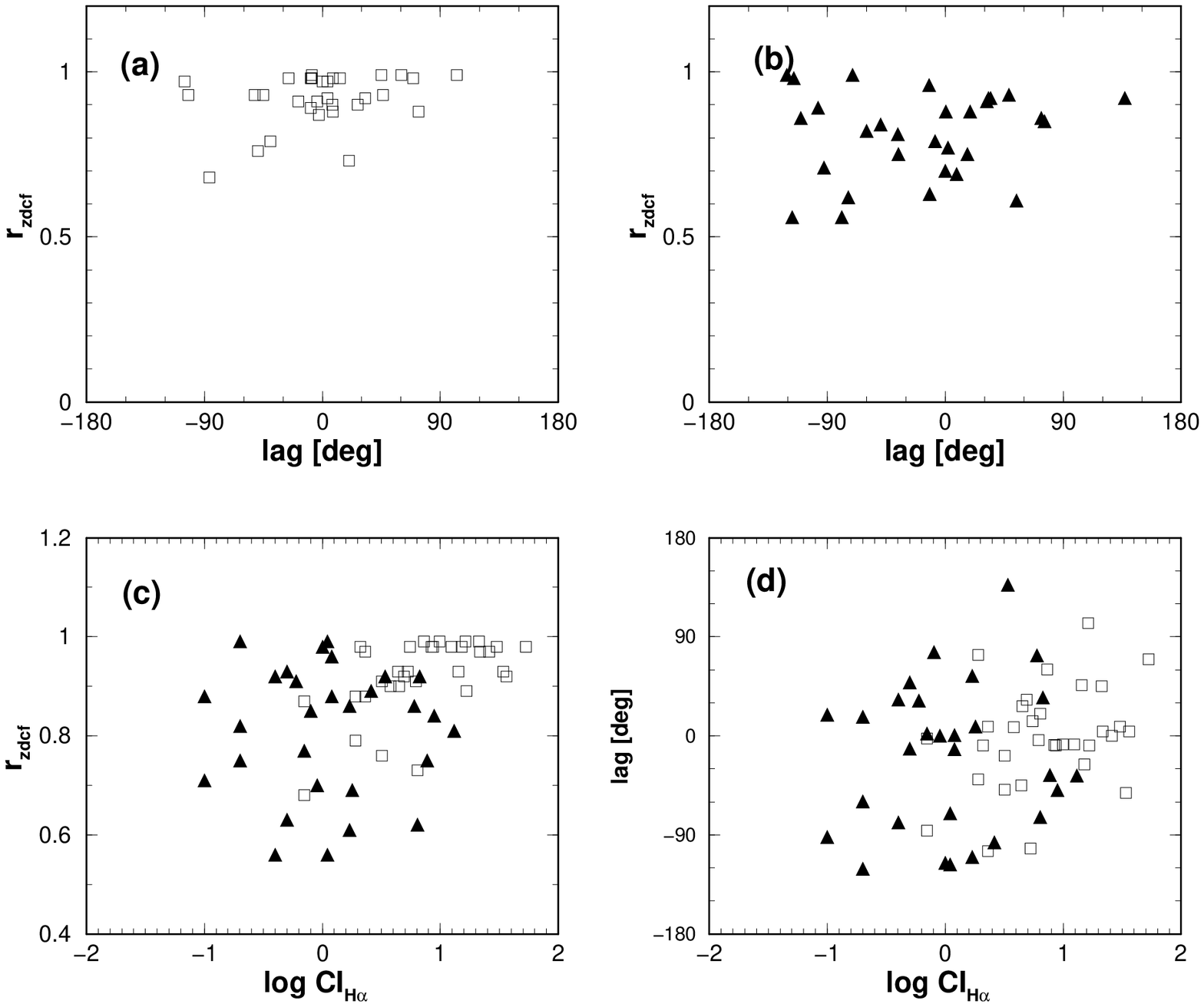}
\figcaption{ Maximum correlation coefficient r$_{zdcf}$ of $H\alpha$ vs.
                     $Cont$,  lag in degrees:\newline
                    BCDs in (a), LSBs in (b),
                    r$_{zdcf}$  vs. log~CI$_{H\alpha}$  in (c), 
                   lag vs. log~CI$_{H\alpha}$  in (d). 
                  Symbols as~for~Fig.~\ref{real2x2}. \label{cross}}
\end{figure}

\newpage
\begin{figure}{}
\vspace{15cm}  
\includegraphics{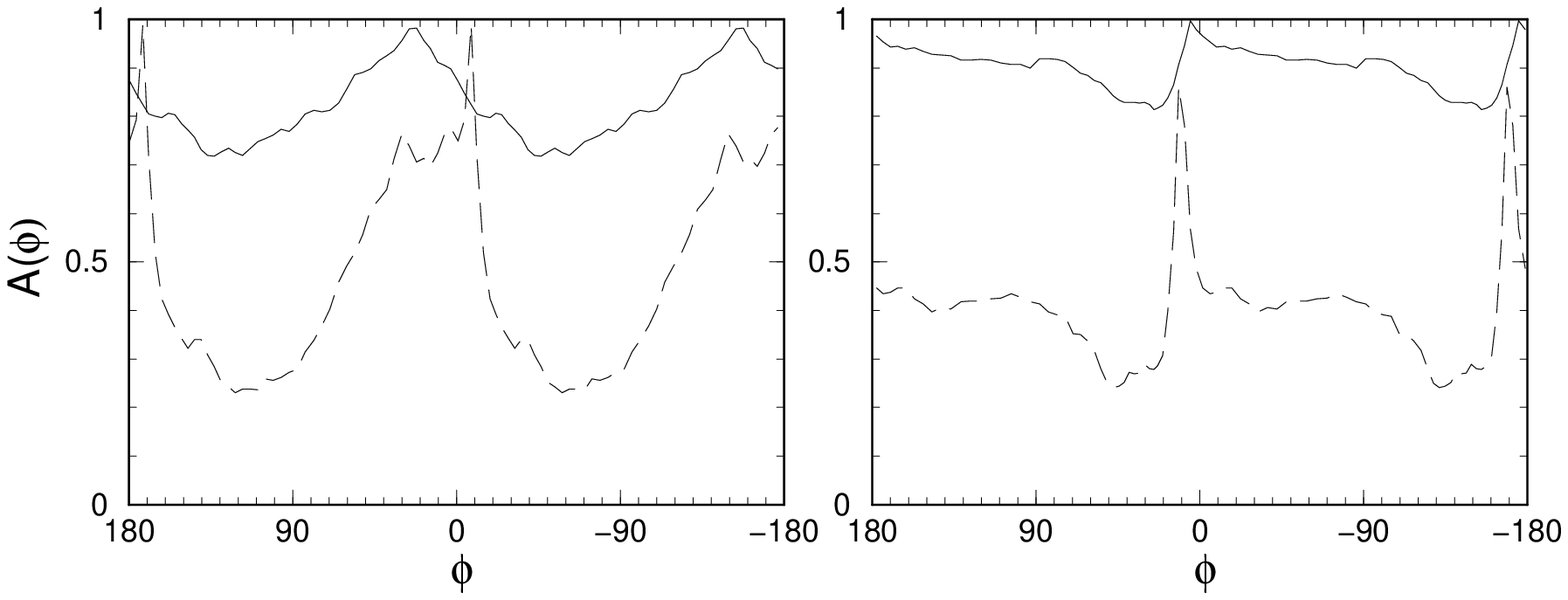}
\includegraphics{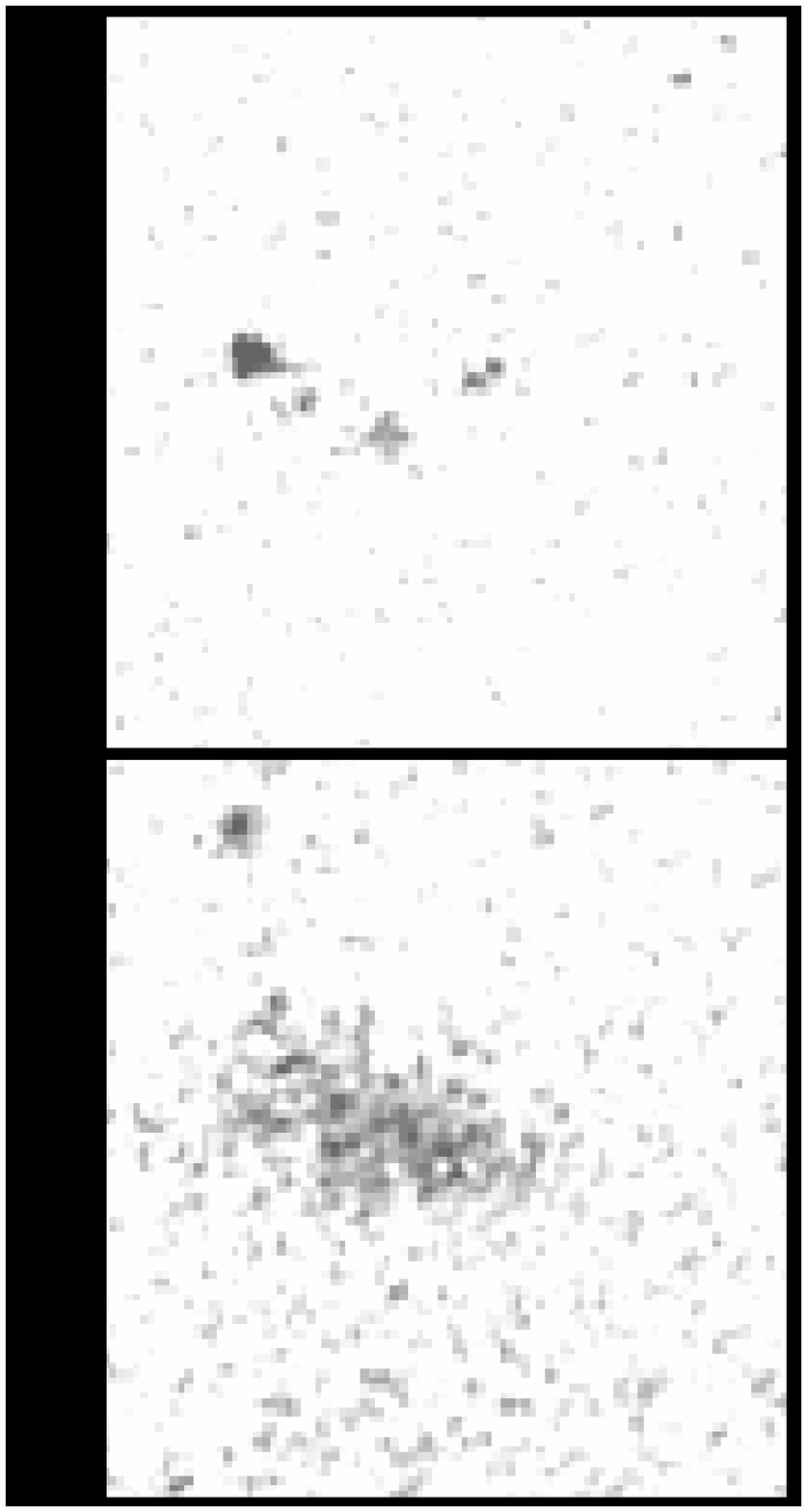}
\includegraphics{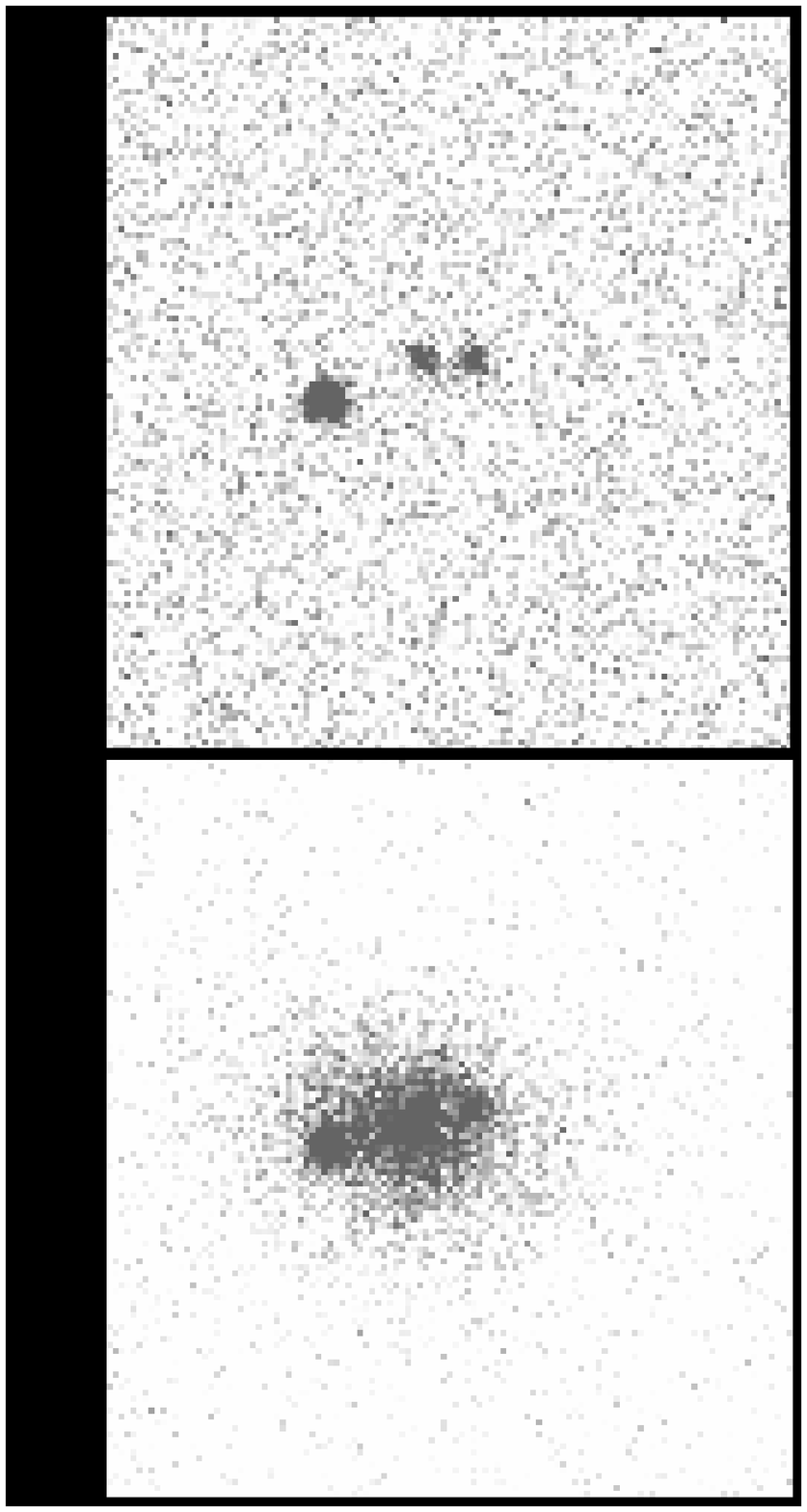}
\figcaption{Comparison of a real galaxy (VCC1753) in the left part of the page
               and a simulated one in the right part.
                          We show in the upper panels the continuum and H$\alpha$
                         images of both galaxies, and in the lower panels 
                          their azimuthal profiles. \label{example}}
\end{figure}

\newpage
\begin{figure}{}
\epsscale{0.75}
\plotone{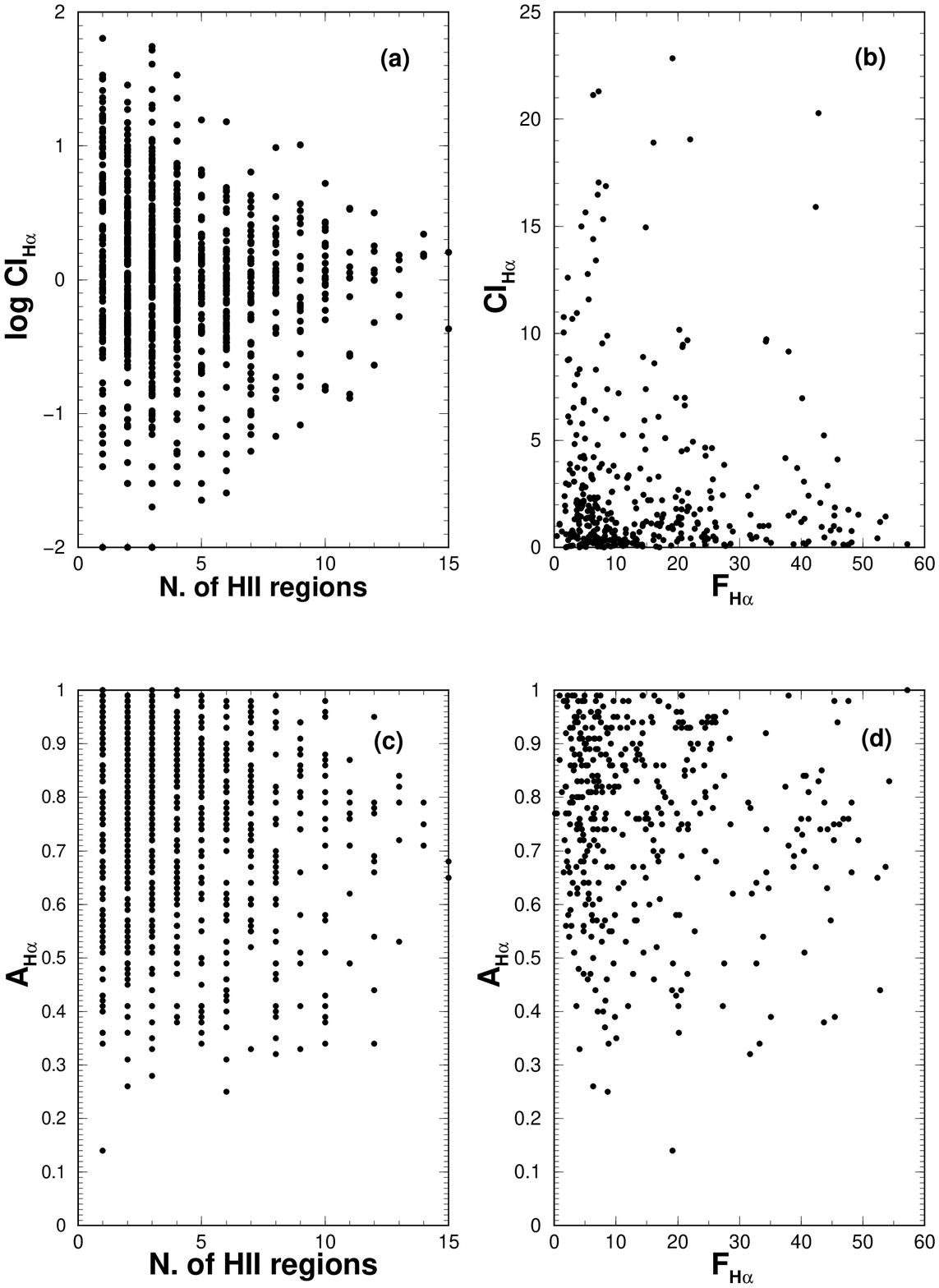}
\figcaption{Artificial galaxies: (a) Concentration index of $H\alpha$ flux vs. 
                        number of HII regions, (b) Concentration index of $H\alpha$ flux vs. 
                    total $H\alpha$ flux, (c) Lopsidedness vs.  number of HII regions, (d) 
                   Lopsidedness vs. total $H\alpha$ flux.  The $H\alpha$ flux
                   is in units of 10$^{-14}$~erg~cm$^{-2}$~s$^{-1}$ \label{model2x2}}
\end{figure}

\newpage
\begin{figure}{}
\plotfiddle{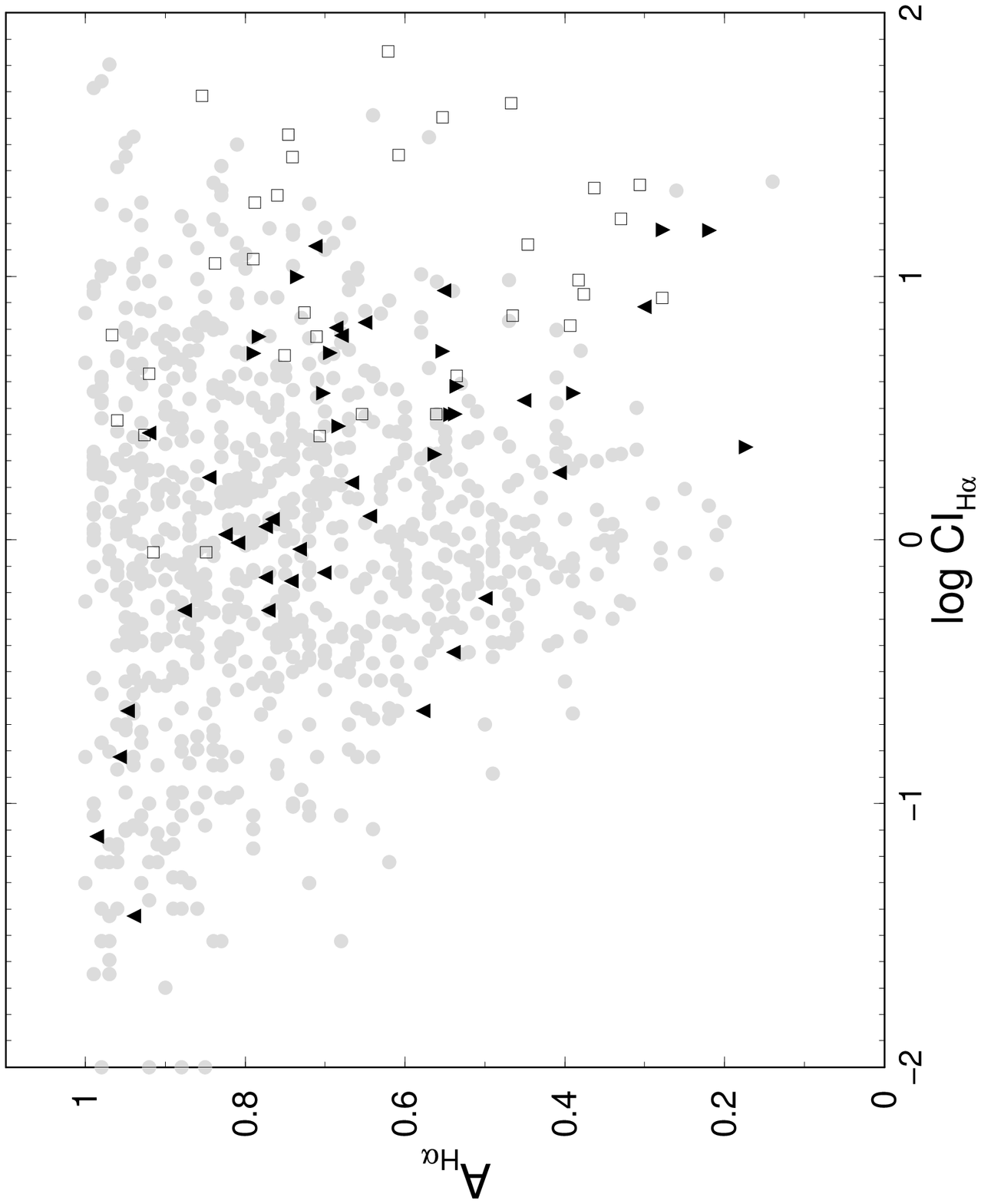}{11cm}{-90}{80}{80}{-260}{450}
\figcaption{A$_{H\alpha}$  vs. log(CI$_{H\alpha}$).   Simulated galaxies  with 1 to 12 
                       HII regions are   represented by  circles,  
                    observed LSBs are  represented by triangles-up for 
                    objects from Heller {\it et al.}~1999    and
                    for  van Zee {\it et al.}~1997a, b (secondary),
                    triangles-down objects from van Zee {\it et al.}~1997a, b    (primary), 
                  observed BCDs are represented by squares. \label{modelall}}
\end{figure}

\newpage
\begin{figure}{}
\plotone{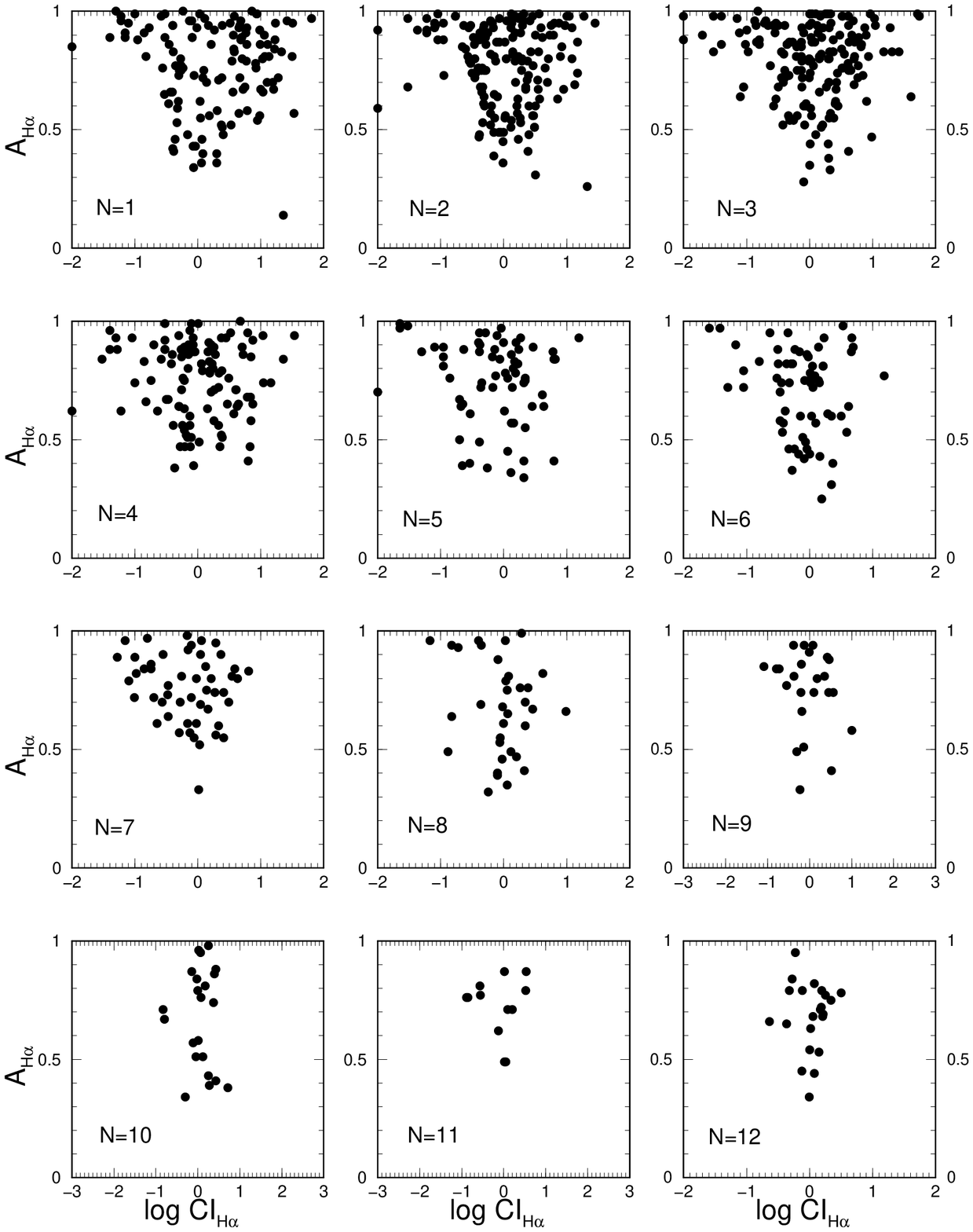}
\figcaption{A$_{H\alpha}$  vs. log(CI$_{H\alpha}$) for simulated galaxies,
                with 1 to 12 HII regions. \label{model12}}
\end{figure}

\newpage
\begin{figure}{}
\plotfiddle{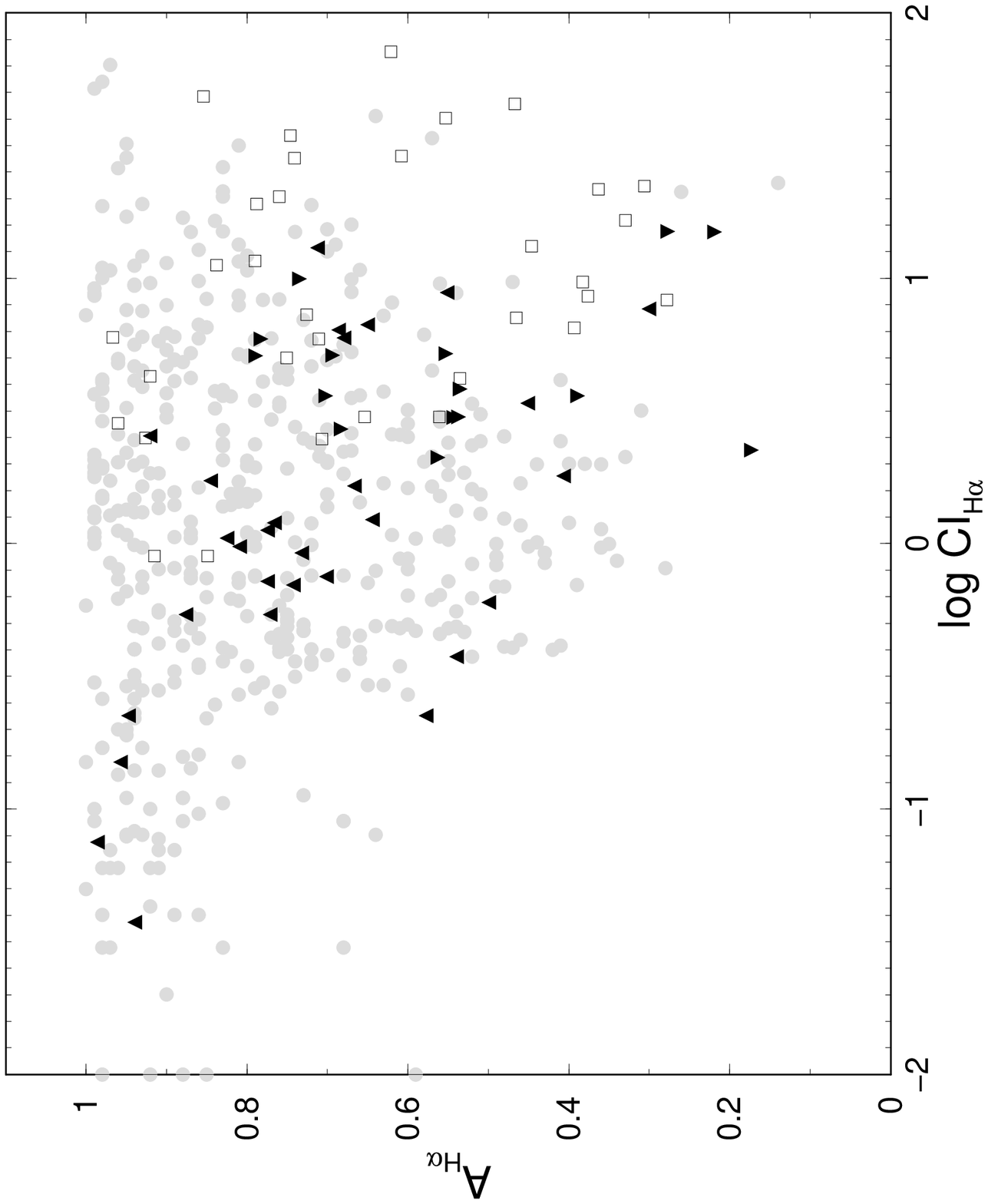}{11cm}{-90}{80}{80}{-260}{450}
\figcaption{Same as Fig.~\ref{modelall}  for simulated galaxies with  1 to 3 HII regions.\label{model123}}
\end{figure}
\end{document}